\documentclass[12pt]{article}
\usepackage{graphicx}
\usepackage{scicite}
\usepackage{bm}
\usepackage{times}
\usepackage{amsmath}
\usepackage{amssymb}
\usepackage{times}
\usepackage{physics}
\usepackage{hyperref}
\usepackage{relsize}
\newenvironment{sciabstract}{%
\begin{quote} \bf}
{\end{quote}}

\begin{document} 
\baselineskip24pt

\title{The Wiedemann-Franz law in doped Mott insulators without quasiparticles}
\author
{Wen O. Wang,$^{1,2,\ast}$ Jixun K. Ding,$^{1,2}$ Yoni
Schattner,$^{2,3,4}$ Edwin W. Huang,$^{5,6,7}$\\ Brian Moritz,$^{2}$ Thomas P. Devereaux,$^{2,8,9,\dagger}$\\
\\
\normalsize{$^{1}$Department of Applied Physics, Stanford University, Stanford, CA 94305, USA}\\
\normalsize{$^{2}$Stanford Institute for Materials and Energy Sciences, 
SLAC National Accelerator Laboratory,}\\ 
\normalsize{2575 Sand Hill Road, Menlo Park, CA 94025, USA}\\
\normalsize{$^{3}$Department of Physics, Stanford University, Stanford, CA 94305, USA}\\ 
\normalsize{$^{4}$AWS Center for Quantum Computing, Pasadena, CA 91125, USA}\\
\normalsize{$^{5}$Department of Physics and Institute of Condensed Matter Theory,}\\ 
\normalsize{University of Illinois at Urbana-Champaign, Urbana, IL 61801, USA}\\
\normalsize{$^{6}$Department of Physics and Astronomy, University of Notre Dame, Notre Dame, IN 46556, USA}\\
\normalsize{$^{7}$Stavropoulos Center for Complex Quantum Matter, University of Notre Dame, Notre Dame, IN 46556, USA}\\
\normalsize{$^{8}$Department of Materials Science and Engineering, Stanford University, Stanford, CA 94305, USA} \\
\normalsize{$^{9}$Geballe Laboratory for Advanced Materials, Stanford University, Stanford, CA 94305, USA} \\
\\
\normalsize{$^\ast$E-mail: wenwang.physics@gmail.com.}
\normalsize{$^\dagger$E-mail: tpd@stanford.edu.}
}

\date{November 30, 2023}

\maketitle 

\begin{sciabstract}
Many metallic quantum materials display anomalous transport phenomena that defy a Fermi liquid description.
Here, we use numerical methods to calculate thermal and charge transport in the doped Hubbard model and observe a cross-over separating high- and low-temperature behaviors. Distinct from the behavior at high temperatures, the Lorenz number $L$ becomes weakly doping dependent and less sensitive to parameters at low temperatures. At the lowest numerically accessible temperatures, $L$ roughly approaches the Wiedemann-Franz constant $L_0$, even in a doped Mott insulator that lacks well-defined quasiparticles. 
Decomposing the energy current operator indicates a compensation between kinetic and potential contributions, which may help to clarify the interpretation of transport experiments beyond Boltzmann theory in strongly correlated metals. 
\end{sciabstract}

Landau's notion of quasiparticles greatly simplified the language of transport in systems with a macroscopic number of interacting degrees of freedom in terms of ``free'' objects with renormalized properties that participate in transport through a semi-classical or Boltzmann framework. 
As such, transport behavior of Fermi liquids is governed solely by kinematic constraints of a Fermi surface and collisions between otherwise free particles. 
Yet in many correlated metals, including the high transition temperature (or critical temperature, $T_c$) cuprates, anomalous transport phenomena have been uncovered that violate these rules:
strange metal resistivity that increases linearly with temperature, not saturating as the quasiparticle mean-free-path approaches the lattice spacing~\cite{RevModPhys.75.1085,hussey2004universality,doi:10.1126/science.abh4273}; 
inconsistency with Kohler's rule, which governs the scaling behavior of magnetoresistance from Boltzmann theory~\cite{PhysRevLett.75.1391,Ayres2021,PhysRevB.53.8733}; 
and violations of the Wiedemann-Franz law, which constrains the ratio of thermal to electrical conductivity~\cite{PhysRevB.49.9073,Hill2001,PhysRevLett.89.147003,PhysRevB.68.220503,PhysRevB.68.100502,PhysRevB.72.214511,PhysRevB.80.104510,PhysRevB.93.064513,zhang2017anomalous,PhysRevX.8.041010,Grissonnanche2019,PhysRevB.100.241114}. 

The ubiquity of such behavior that violates notions of the Fermi liquid has led to tremendous interest in determining how heat and charge currents propagate in systems without the saving grace of quasiparticles~\cite{hartnoll2015theory,RevModPhys.94.041002,PhysRevB.88.125107,PhysRevLett.119.141601,PhysRevB.106.245123}.
Analysis of the large body of experimental transport results in correlated materials has been hindered dramatically by the use of an assumed Boltzmann-like theory and reductive conclusions on the nature of transport in terms of Drude-like single-particle concepts.
While this greatly amplifies the need for deeper analysis that avoids oversimplifications, there is very little known from exact methods about the nature of transport in strongly interacting systems. 
Many advanced numerical calculations have focused on characterizing ground state properties~\cite{doi:10.1146/annurev-conmatphys-031620-102024,doi:10.1146/annurev-conmatphys-090921-033948},
but a picture of transport is incomplete without an understanding of the excited states in these materials. 
Analytical approaches are hampered by the fact that properly evaluating transport involves calculating many higher order correlation functions without relying on the simplifying assumptions of quasiparticles and Boltzmann theory, which only punctuates the need for more accurate and precise determinations of transport. 

Here, we numerically study the DC longitudinal thermal conductivity $\kappa$ in the doped two-dimensional (2D) $t$-$t'$-$U$ Hubbard model, which exhibits strange metallic electric transport over a wide hole doping $p$ and temperature $T$ range~\cite{edwin,brown2019bad,nichols2019spin,xu2019bad}.
We evaluate the many-body Kubo formula using the determinant quantum Monte Carlo (DQMC)~\cite{DQMC1,DQMC2} algorithm, which is numerically exact, unbiased, and non-perturbative,
and maximum entropy analytic continuation (MaxEnt)~\cite{jarrell1996bayesian,PhysRevB.82.165125}, which is typically reliable in systems with strong interactions that lack sharp features in frequency [see supplementary materials of~\cite{edwin}]
We define $\kappa$ as the linear response of the heat current $\langle\mathbf{J}_Q\rangle$ induced by a parallel temperature gradient and normalized by system size $N$,
$\kappa\equiv-\langle J_{Q,x}\rangle/(N\partial_x T)$, under the condition of zero charge current.
Distinct from the incoherent behavior at high temperatures, we observe that the Lorenz number, the ratio between the thermal and charge conductivity $L\equiv \kappa/(T\sigma)$, has a weak doping and parameter dependence in the low-temperature regime and roughly approaches the Wiedemann-Franz law prediction $L_0=\pi^{2}/3$ as temperature decreases down to the lowest accessible value, even in the absence of long-lived quasiparticles.
Methodological details, including a systematic analysis of finite size and Trotter errors, as well as extensive supporting data, can be found in~\cite{supplementary}.

\section*{Thermal and charge conductivity}

\begin{figure}
    \centering
    \includegraphics[width=\textwidth]{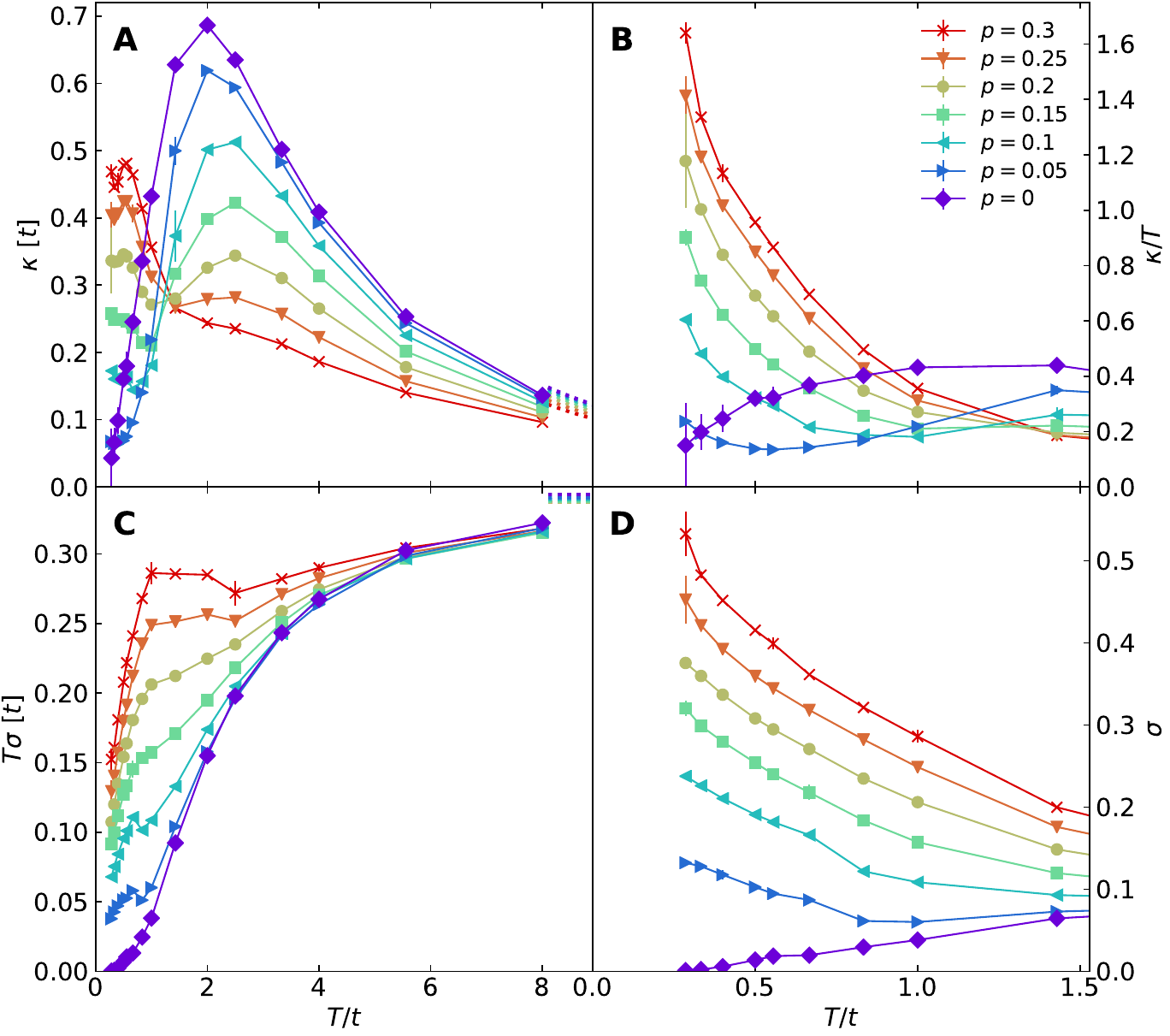}
    \caption{
    \textbf{Temperature and doping dependence of thermal and charge conductivity.}
    (\textbf{A}) DC thermal conductivity $\kappa$.
    (\textbf{B}) $\kappa/T$ focused on the low-temperature regime.
    (\textbf{C}) DC charge conductivity $\sigma$ multiplied by temperature $T$.
    (\textbf{D}) $\sigma$ focused on the low-temperature regime.
    The high-temperature dotted lines in (A) and (C) are infinite-temperature limits calculated via a moments expansion ~\cite{PhysRevB.105.L161103,edwin}.
    Parameters: $U/t=8$ and $t'/t=-0.25$.
    A cross-over temperature $T_{\mathrm{xo}}\sim t$ separates low- and high-temperature regimes in (A) and (C). Error bars are shown but may be smaller than the size of the data markers.
    }
    \label{fig:conductivities}
\end{figure}

The DC longitudinal thermal conductivity $\kappa(T)$ is shown in Fig.~\ref{fig:conductivities}A; 
for comparison, the DC longitudinal charge conductivity $\sigma(T)$~\cite{edwin} (multiplied by $T$) is shown in Fig.~\ref{fig:conductivities}C.
In the infinite-temperature limit, $\kappa \propto 1/T^2$ and $\sigma \propto 1/T$~\cite{edwin,PhysRevB.105.L161103,brown2019bad,Huse}.
As $T$ decreases from this limit, we observe a cross-over at roughly $T_{\mathrm{xo}} \sim t$, 
separating distinct behavior in two regimes for both $\kappa$ and $\sigma$.
$\kappa$ decreases with doping at high temperatures, whereas it increases with doping at low temperatures.
Although $\sigma$ generally increases with doping at all temperatures, the temperature dependence of $T\sigma$ displays kinks, or even nonmonotonic behavior, at roughly $T_{\mathrm{xo}}$.
Below $T_{\mathrm{xo}}$, $\kappa/T$ and $\sigma$ display similar doping and temperature dependences (Fig.~\ref{fig:conductivities}, B and D), suggesting persistent correlations between thermal and charge transport even for a strange metal phase where quasiparticles are not well-defined.

\section*{Lorenz number and its temperature and parameter dependence}

The Lorenz number $L(T)$ highlights the correlation between thermal and charge transport (Fig.~\ref{fig:lorenz}).
Aside from the half-filled Mott insulator, where $L$ diverges with decreasing temperature, in the doped metals $L$ shows a cross-over similar to that in $\kappa$ and $\sigma$. At high temperatures, high-energy excited states become important~\cite{Huse,Kadanoff}, such that quasiparticles are not well-defined and electrons have extraordinarily short mean-free-paths. $L$ has a nonmonotonic temperature dependence and decreases with increasing doping. 
Below $T_{\mathrm{xo}}$, $L$ displays substantially reduced doping dependence, collapsing roughly onto a single set of curves. This set of curves increases monotonically with decreasing temperatures, approaching a constant that roughly corresponds to $L_0= \pi^2/3$ -- the Lorenz number as predicted by the Wiedemann-Franz law.

In the Hubbard model, relaxation primarily occurs through Umklapp scattering. To test its impact on the conductivities and $L$, we modulate Umklapp scattering by modifying the Hubbard $U$ and next-nearest-neighbor hopping $t'$, with the results shown in Fig.~\ref{fig:parameters}.
The high-temperature peak position of $L$ is largely controlled by $U$, increasing with increasing $U$, similar to the behavior of the specific heat [see fig.~S9 in~\cite{supplementary}]. 
For temperatures below the cross-over,
there is no strong dependence of $L$ on either $U$ or $t'$, suggesting that the low-temperature behavior is quite generic to the strongly correlated Hubbard model: changing the shape of the Fermi surface ($t'$) or the strength of Umklapp scattering ($U$) does not appreciably alter $L$ at the temperatures accessible through DQMC.

\section*{Decomposing the Lorenz number}

To better understand the behavior below $T_{\mathrm{xo}}$, it is useful to look at the operator contributions to the conductivities.
Determining $\kappa$ in the Hubbard model using the Kubo formula requires one to consider the two-particle term in the energy current operator arising from electron-electron interactions, as opposed to Boltzmann theory that relies entirely on single-particle properties.
The energy current operator $\mathbf{J}_{E}$ consists of a single-particle kinetic energy contribution $\mathbf{J}_{K}$, similar to that appearing in the charge current operator $\mathbf{J}$, plus an additional term $\mathbf{J}_{P}$, which we call the potential energy current that depends explicitly on the interaction and importantly contains 
a two-particle current [see Eq. S2, Eq. S3, and the relevant discussion of the Formalism in~\cite{supplementary}]. The heat current $\mathbf{J}_Q$, from which we obtain $\kappa$, itself contains an additional term $-\mu\mathbf{J}$, where $\mu$ is the chemical potential. However, under the condition of zero charge current $\left(\langle\mathbf{J}\rangle=0\right)$, terms proportional to $\langle\mathbf{J}\rangle$ will not contribute to $\langle\mathbf{J}_Q\rangle$, leaving only $\langle\mathbf{J}_{K}\rangle$ and $\langle\mathbf{J}_{P}\rangle$.
In this way, we separate $\kappa$ into kinetic and potential contributions $\kappa_{K/P}\equiv-\langle J_{K/P,x}\rangle/(N\partial_x T)$. Similarly, we can express the Lorenz number $L$ as a sum of its kinetic and potential contributions, with $L = L_K + L_P$, where $L_{K/P} \equiv \kappa_{K/P}/(T\sigma)$ (Fig.~\ref{fig:KPsep}, A and B).

\begin{figure}
    \centering
    \includegraphics[width=\textwidth]{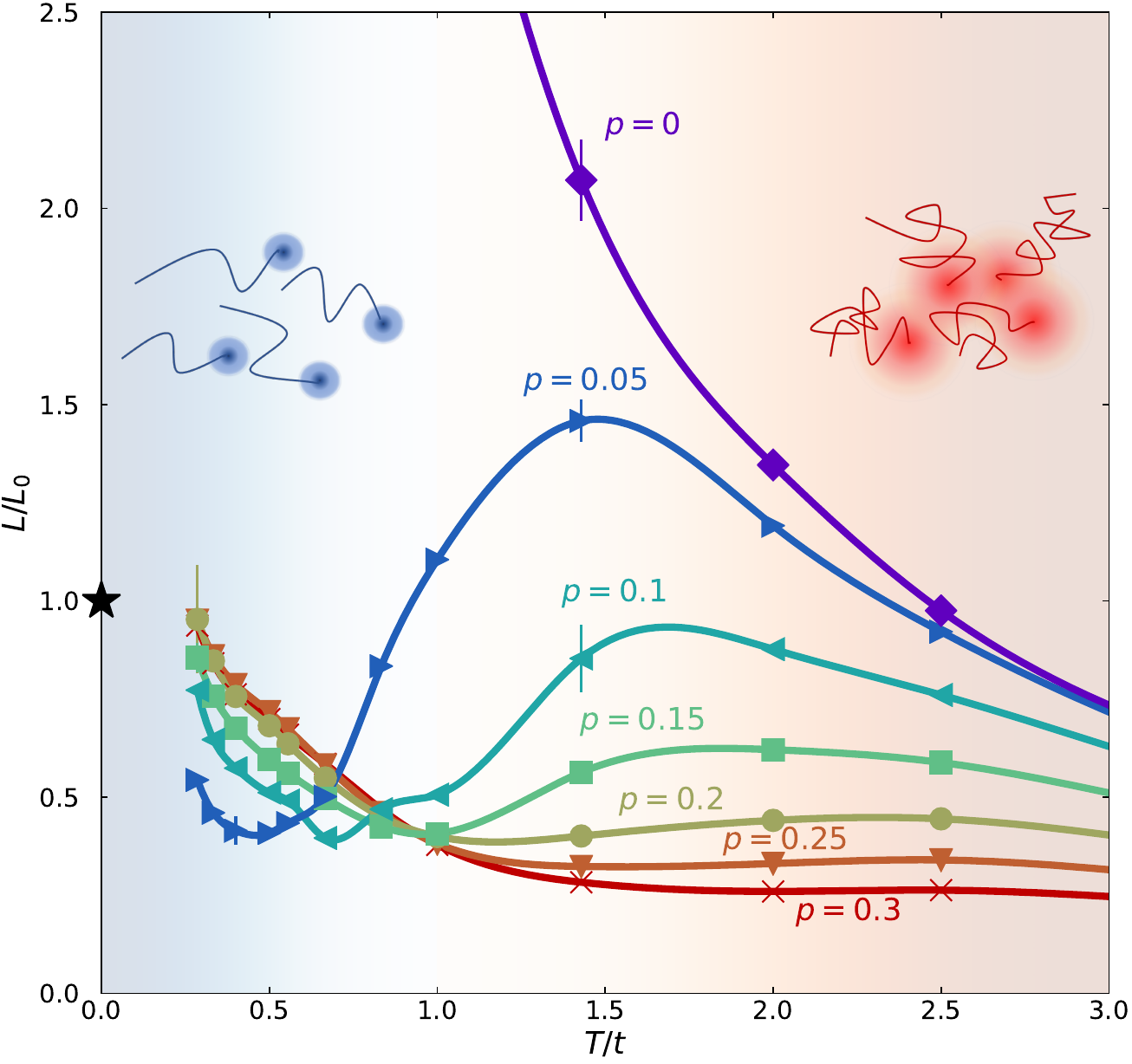}
     \caption{
    \textbf{Lorenz number}. $L=\kappa/(T\sigma)$ normalized by $L_0$.
    The lines are guides to the eye. At low temperatures, below $T_{\mathrm{xo}}\sim t$, $L/L_0$ approaches roughly $1$, marked by the black star. 
    Parameters: $U/t=8$ and $t'/t=-0.25$.
    Cartoons: at high temperatures, high-energy excited states are important~\cite{Huse,Kadanoff} and transport is incoherent; electrons are strongly correlated and have an extraordinarily short mean-free-path. 
    At low temperatures, the electrons are on their way toward some sort of ``coherence''; electrons have a longer mean-free-path, although not long enough for well-defined long-lived quasiparticles. While single-particle and individual transport properties show signatures of anomalous strange metal and non-Fermi liquid behavior, the Lorenz number still roughly approaches the Wiedemann-Franz law's prediction as temperature decreases.
    }
    \label{fig:lorenz}
\end{figure}

\begin{figure}
    \centering
    \includegraphics[width=\textwidth]{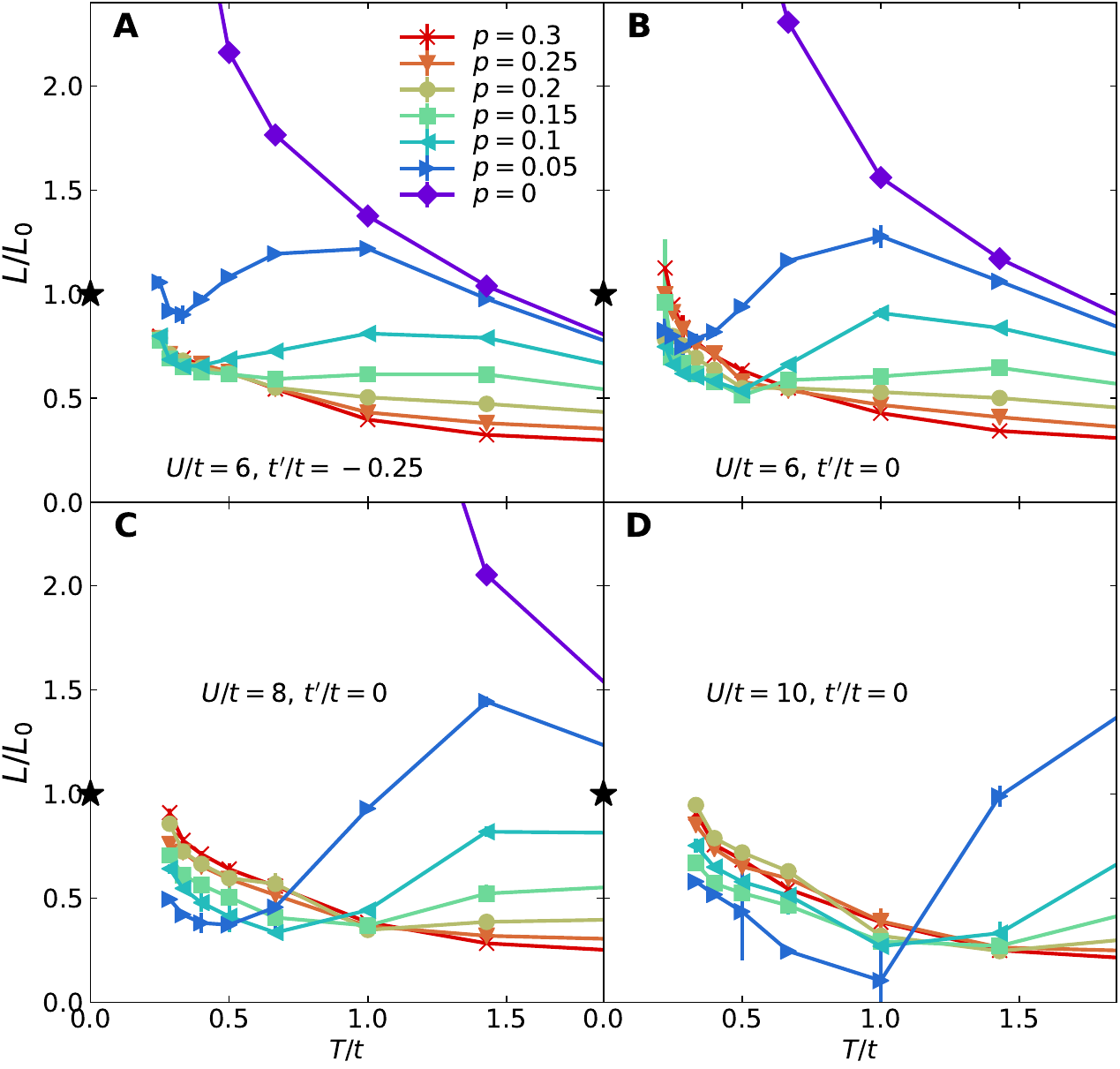}
    \caption{
     \textbf{Parameter dependence of the Lorenz number}.
    $L/L_0$ for (\textbf{A}) $U/t=6$ and $t'/t=-0.25$;
    (\textbf{B}) $U/t=6$ and $t'/t=0$;
    (\textbf{C}) $U/t=8$ and $t'/t=0$;
    (\textbf{D}) $U/t=10$ and $t'/t=0$.
   The black stars mark the value $1$.
   The lowest temperatures are lower for smaller $U$ due to a better behaved fermion sign.
 }
    \label{fig:parameters}
\end{figure}

\begin{figure}
    \centering
    \includegraphics[width=\textwidth]{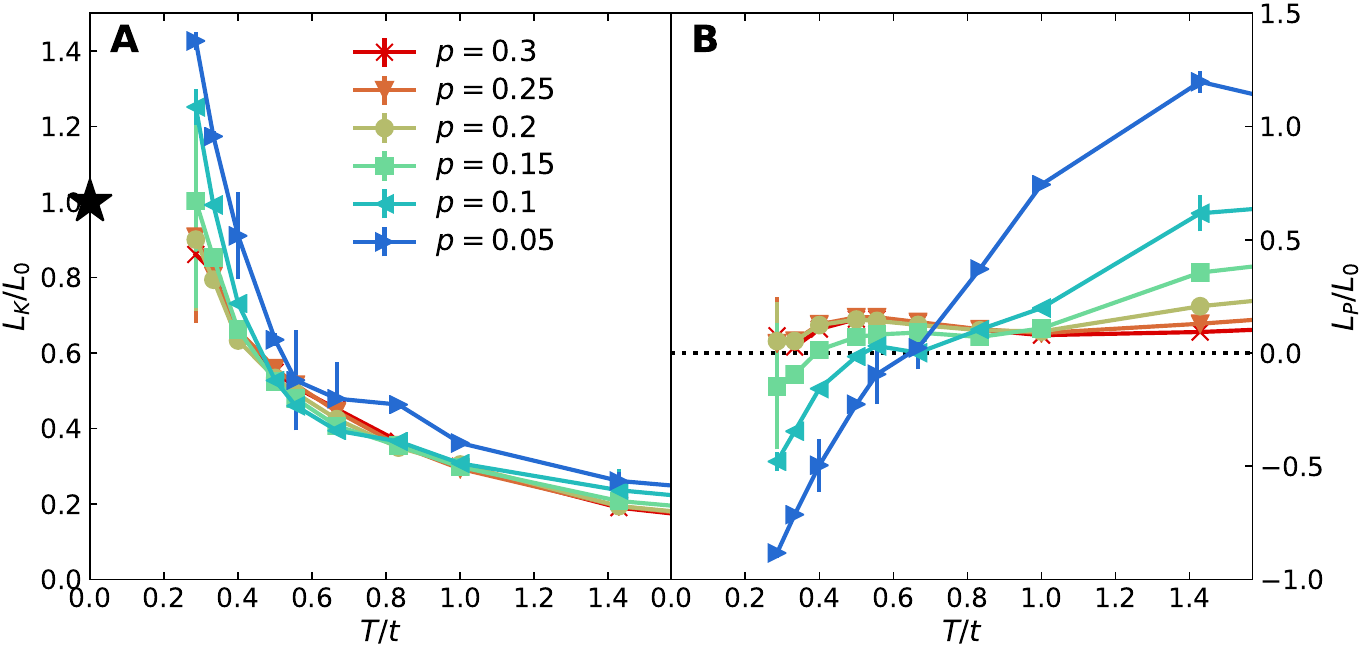}
    \caption{
     \textbf{Kinetic and potential decomposition of the Lorenz number}.
    (\textbf{A}) Normalized kinetic contribution $L_K/L_0$.
   The black star marks the value $1$.
    (\textbf{B}) Normalized potential contribution $L_P/L_0$.
    The black dotted line marks the value $0$.
    Parameters: $U/t=8$ and $t'/t=-0.25$.
 }
    \label{fig:KPsep}
\end{figure}

At high temperatures, the kinetic energy contribution $L_K$ is relatively small and doping independent, while the potential energy contribution $L_P$ is large at small doping and decreases for increasing hole concentration due to the reduction of double occupancies. This doping dependence is imparted to the combined $L$ (as already shown in Fig.~\ref{fig:lorenz}). Below the cross-over temperature $T_{\mathrm{xo}}$ and at large doping, $L_P$ is relatively small and $L_K$ and $L$ approach $L_0$. At low doping, $L_K$ increases with decreasing temperature, while $L_P$ decreases and changes sign at roughly $T_{\mathrm{xo}}$.
The separate contributions from the kinetic and potential terms show opposing behavior, which becomes more dramatic for lower doping, and effectively compensate one another, resulting in $L$ that approaches $L_0$.
Thus unexpectedly, the ratio $L$ displays a relative insensitivity to doping, and Hubbard model parameters [see fig.~S10 in~\cite{supplementary}], at the lowest accessible temperatures.

\section*{Discussion and outlook}

The congruence between charge and thermal transport in the Hubbard model is surprising.
For scattering dominated by elastic processes, such as disorder or quasi-elastic phonon scattering above the Debye temperature, the thermal and charge conductivity are correlated through the Wiedemann-Franz law~\cite{Mousatov2021,PhysRevB.80.104510,PhysRevB.88.125107,tulipman2022criterion}, such that for $T$ much lower than the Fermi temperature, one obtains the Lorenz number $L=L_0=\pi^2/3$.
For both Fermi liquids and non-Fermi liquids without disorder, $L$ deviates substantially from $L_0$~\cite{tulipman2022criterion,PhysRevB.88.125107,PhysRevB.99.085104}. 
Despite our lack of knowledge about the exact behavior of the Hubbard model at lower temperatures (Fermi liquid or not) due to the fermion sign problem, the result that $L$ approaches a weakly doping and Hubbard parameter dependent constant very close to $L_0$ indicates a surprisingly universal behavior. This behavior is observed only when both single- and two-particle contributions are properly accounted for in the heat-current operator. 

Our results may be understood in three possible ways. First, although the temperatures in our study are below the magnetic exchange energy scale $J$, our results may not yet be in the asymptotic low temperature regime to assess the $T\rightarrow 0$ limit. Second, one might expect the approximate Wiedemann-Franz ratio to emerge in a system where both charge and thermal currents relax predominantly through Umklapp scattering in our temperature regime. Lastly, it may be that such a compensation effect between kinetic and potential energy contributions to $L$ cannot be cast in the usual Boltzmann like formulation for strongly interacting, anisotropic systems such as the Hubbard model. 

Finally, what can our results say about the strong violation of the Wiedemann-Franz law that has been observed in cuprates at room temperature, with $L$ larger than $L_0$ by a factor of $3$ or more~\cite{Mousatov2021,PhysRevB.49.9073,PhysRevB.68.220503,PhysRevB.100.241114}? 
One explanation for this is that the strong interaction enhances the electronic contribution to thermal transport, while another explanation would rely on a substantial phonon contribution to the heat current. 
Our observation over the experimentally relevant temperature range that the electronic contribution $L$ roughly approaches $L_0$ from below would be consistent with scenarios in which the large $L$ in cuprates requires an appreciable phonon contribution to heat transport.

\bibliographystyle{Science}

\section*{Acknowledgments}

We acknowledge helpful discussions with A. Auerbach, D. Belitz, E. Berg, S. A. Hartnoll, N. E. Hussey, S. A. Kivelson, P. A. Lee, R. T. Scalettar, Z. X. Shen, and E. Tulipman. 

\section*{Funding}
This work was supported by the US Department of Energy (DOE), Office of Basic Energy Sciences,
Division of Materials Sciences and Engineering. 
E.W.H. was supported by the Gordon and Betty Moore Foundation EPiQS Initiative through the grants GBMF 4305 and GBMF 8691.
Y.S. was supported by the  Gordon and Betty Moore Foundation’s EPiQS Initiative through grants  GBMF 4302 and GBMF 8686.
Y.S.'s contribution to the work was done prior to joining AWS Center for Quantum Computing.
Computational work was performed on the Sherlock cluster at Stanford University and on resources of the National Energy Research Scientific Computing Center, supported by the U.S. DOE, Office of Science, under Contract no. DE-AC02-05CH11231.

\section*{Author contributions}
W.O.W. and T.P.D. conceived the study. W.O.W. performed numerical simulations and conducted data analysis and interpretation. J.K.D., Y.S., E.W.H., B.M., and T.P.D. assisted in data interpretation. W.O.W., B.M., and T.P.D. wrote the manuscript with input from all coauthors.

\section*{Competing interests}
The authors declare no competing interests.

\section*{Data and materials availability}
Code and data presented in this study are deposited in Zenodo (41, 42).

\clearpage

\clearpage
\baselineskip14.5pt

\renewcommand{\thefigure}{S\arabic{figure}}
\setcounter{figure}{0}
\renewcommand{\theequation}{S\arabic{equation}}
\setcounter{section}{0}

\section*{Supplementary Materials}
Methods\\
Supplementary Text\\
Figs.~S1 to S15\\
References (\textit{43-47})
\setcounter{page}{1}

\clearpage

\subsection*{Methods}

\noindent\underline{Simulation Parameters}

Simulations of the 2D single-band $t$-$t'$-$U$ Hubbard model in the grand canonical ensemble were performed on $8\times 8$ square lattice clusters (unless otherwise specified), with periodic boundary conditions, next-nearest-neighbour hopping $t'=-0.25t$ or $0$, and on-site Coulomb repulsion $U$ between $6t$ and $10t$. 
For convenience, $k_B$, $\hbar$, and charge $e$ are set to $1$ throughout the paper.
Measurements of all quantities, other than particle density $\ev{n}$ in the chemical potential tuning process described later, if not otherwise specified, were performed with an imaginary time Trotter discretization $d\tau=0.05/t$~\cite{DQMC1,DQMC2}. At high temperatures, the smallest number of imaginary time slices used in the Trotter decomposition was $\tilde{L}=\beta/d\tau = 20$, where $\beta$ is the inverse temperature $T^{-1}$.
Each Markov chain in the Monte Carlo process consisted of $5\times 10^4$ warm up sweeps and $10^6$ measurement sweeps through space-time. Unequal time measurements were taken every $4$ sweeps. Measurements were made on up to $\sim 2400$ Markov chains for each set of parameters at the lowest temperatures, with $2.5\times 10^5$ unequal time measurements and approximately $2\times 10^5 \times \tilde{L}$ equal time measurements per chain.%

\noindent\underline{Analytic Continuation}

We evaluated DC transport coefficients using the Kubo formula for $\kappa$ and $\sigma$~\cite{PhysRevB.105.L161103}. 
The DC transport coefficients were obtained by performing maximum entropy analytic continuation (MaxEnt)~\cite{jarrell1996bayesian,PhysRevB.82.165125} on DQMC measurements of corresponding correlation functions in imaginary time.
To determine the adjustable parameter which assigns weights of statistics and entropy in the maximized function in MaxEnt, we use the method of Ref.~\cite{PhysRevE.94.023303}.
Details of the formalism, operators, and specific correlation functions can be found in the ``Formalism'' subsection of the Supplementary Text.

The MaxEnt algorithm requires a ``model'' function to regularize the real-frequency correlation function. In this work, we used an annealing procedure in which spectra from one temperature serve as the model function for the next lower temperature in a sequence~\cite{PhysRevB.105.L161103,edwin}.
We determined spectra in the infinite-temperature limit using the moments expansion method~\cite{PhysRevB.105.L161103,edwin} (up to sixth order for $t'/t=-0.25$, or eighth order for $t'=0$), and used these spectra as the model functions at the highest temperature. 
Although the choice of model function may impact the result of the MaxEnt analysis, this variation does not significantly affect the quantitative results and the qualitative behavior and conclusions remain unchanged (see Fig.~\ref{fig:differentana} in the Supplementary Text). 

\noindent\underline{Chemical Potential Tuning}

To tune the chemical potential $\mu$ for a specific target filling $n_{\mathrm{tar}}=1-p_{\mathrm{tar}}$ at a given temperature and lattice size, we used DQMC to calculate $\langle n \rangle$ over a range of chemical potentials $\mu$ (at $0.05t$ intervals). We obtained the best $\mu$ by interpolation of $\langle n \rangle$ versus $\mu$. For the tuning process, the maximum imaginary time discretization $d\tau$ was chosen to be $0.02/t$, and at high temperatures, the smallest number of imaginary-time slices was $\tilde{L} = 20$. The doping $p$ in each figure indicates the target doping $p_{\mathrm{tar}}$.

\clearpage
\noindent\underline{Error Analysis}

Error bars are shown for all measurements. If not otherwise specified, error bars are determined by bootstrap resampling~\cite{bootstrap}. In particular, we calculated $100$ bootstraps and used the standard deviation of the distribution as the standard error of the mean. The mean values represent the average values from bootstrap resampling. 
For analytic continuation, the average spectra from bootstrap resampling at one temperature served as the model function for the next lower temperature, as described above for annealing. 

\subsection*{Supplementary Text}

\noindent\underline{Formalism} \label{sec:formulism}

In this paper, $\kappa$ refers to the longitudinal DC thermal conductivity measured under the condition of zero charge current~\cite{PhysRevB.105.L161103}, distinguished from the one measured under the condition of zero electric field; $c_v$ refers to the specific heat, defined as the energy density difference of the system per temperature difference at fixed density.

We investigate the 2D single-band Hubbard model with spin $S=1/2$.
Considering both nearest-neighbour $t$ and the next-nearest-neighbour $t'$ hopping, the Hamiltonian is
\begin{eqnarray}
\hat{H} &= - t \mathlarger{\sum}\limits_{\langle lm \rangle \sigma} \left(c^\dagger_{l\sigma}c_{m\sigma}
 + c^\dagger_{m\sigma}c_{l\sigma}\right)  
  - t' \mathlarger{\sum}\limits_{\langle \langle lm \rangle \rangle \sigma} \left(c^\dagger_{l\sigma}c_{m\sigma}
 + c^\dagger_{m\sigma}c_{l\sigma}\right)   \nonumber \\
 &+ U\mathlarger{\sum}\limits_{l} \left(n_{l\uparrow}-\frac{1}{2}\right)\left(n_{l\downarrow}-\frac{1}{2}\right),
\label{hubbard}
\end{eqnarray}
where $U$ is the on-site Coulomb interaction,
$\mathit{c}_{l,\mathit{\sigma}}^{\dagger}$ $(\mathit{c}_{l,\mathit{\sigma}})$ is the creation (annihilation) operator for an electron at site $l$ with spin $\mathit{\sigma}$, and $\mathit{n}_{l,\mathit{\sigma}} \equiv \mathit{c}_{l,\mathit{\sigma}}^{\dagger} \mathit{c}_{l,\mathit{\sigma}}$ is the number operator at site $l$ with spin $\mathit{\sigma}$. 

Expressions for $\kappa$, $\sigma$, and $c_v$ have been derived in the Supplementary Material in Ref.~\cite{PhysRevB.105.L161103}. 
Calculating $\sigma$ and $\kappa$ requires correlation functions that include the particle/charge current $\mathbf{J}$ and energy current $\mathbf{J}_E$ terms. From the Hamiltonian in Eq.~\ref{hubbard}, we derive $\mathbf{J}$ and $\mathbf{J}_E$ in a manner similar to that in Ref.~\cite{PhysRevB.105.L161103}, obtaining
\begin{eqnarray}
 \mathbf{J}
 &=\frac{t}{2}\mathlarger{\sum}\limits_{l,\bm{\delta} \in \mathrm{NN},\sigma}\bm{\delta}
(ic^\dagger_{l+\delta,\sigma}c_{l,\sigma}+h.c.)  \nonumber \\
&+\frac{t'}{2}\mathlarger{\sum}\limits_{l,\bm{\delta}' \in \mathrm{NNN},\sigma}\bm{\delta}'
(ic^\dagger_{l+\delta ',\sigma}c_{l,\sigma}+h.c.) \label{j}
\end{eqnarray}
and
\begin{eqnarray}
\mathbf{J}_E 
&= \mathlarger{\sum}\limits_{l,\bm{\delta}_1 \in \mathrm{NN},\bm{\delta}_2 \in \mathrm{NN},\sigma}(-\frac{\bm{\delta}_1+\bm{\delta}_2}{4}) {t^2}(i c^{\dagger}_{l+\delta_1+\delta_2,\sigma}c_{l,\sigma} + h.c.) \nonumber\\
&+ \mathlarger{\sum}\limits_{l,\bm{\delta}\in \mathrm{NN},\bm{\delta} ' \in \mathrm{NNN},\sigma }(-\frac{\bm{\delta}+\bm{\delta} '}{2})tt'(i c^{\dagger}_{l+\delta+\delta ',\sigma}c_{l,\sigma} + h.c.) \nonumber\\
&+ \mathlarger{\sum}\limits_{l,\bm{\delta}'_1 \in \mathrm{NNN},\bm{\delta}'_2 \in \mathrm{NNN},\sigma}(-\frac{\bm{\delta}'_1+\bm{\delta}'_2}{4}) {t'^2} (i c^{\dagger}_{l+\delta '_1+\delta '_2,\sigma}c_{l,\sigma} + h.c.) \nonumber\\
&+\frac{Ut}{4} \mathlarger{\sum}\limits_{l,\bm{\delta} \in \mathrm{NN},\sigma} \bm{\delta} (n_{l+\delta,-\sigma}+n_{l,-\sigma})(ic_{l+\delta,\sigma}^\dagger c_{l,\sigma}+h.c.) \nonumber\\
&+\frac{Ut'}{4}\mathlarger{\sum}\limits_{l,\bm{\delta} ' \in \mathrm{NNN},\sigma} \bm{\delta} ' (n_{l+\delta ',-\sigma}+n_{l,-\sigma})(ic_{l+\delta ',\sigma}^\dagger c_{l,\sigma}+h.c.) \nonumber\\
&- \frac{Ut}{4}\mathlarger{\sum}\limits_{l,\bm{\delta} \in \mathrm{NN},\sigma}\bm{\delta}
(ic^\dagger_{l+\delta,\sigma}c_{l,\sigma}+h.c.) \nonumber\\
&-\frac{Ut'}{4}\mathlarger{\sum}\limits_{l,\bm{\delta} '\in \mathrm{NNN},\sigma}\bm{\delta} '
(ic^\dagger_{l+\delta ',\sigma}c_{l,\sigma}+h.c.). \label{je}
\end{eqnarray}
To make the notations above clear, $\mathrm{NN}$ ($\mathrm{NNN}$) denotes the set of nearest-neighbour (next-nearest-neighbour) position displacements. Specifically, on the two-dimensional square lattice, $\mathrm{NN}=\{+\mathbf{x}, -\mathbf{x}, +\mathbf{y}, -\mathbf{y}\}$ and $\mathrm{NNN}=\{+\mathbf{x}+\mathbf{y}, -\mathbf{x}+\mathbf{y}, +\mathbf{x}-\mathbf{y}, -\mathbf{x}-\mathbf{y}\}$, where the lattice constant is set to $1$ and $\mathbf{x}$ and $\mathbf{y}$ are unit vectors.
Here, if $l$ is an arbitrary site label associated with the position vector $\mathbf{R}_l = x_l \mathbf{x} + y_l \mathbf{y}$, and $\bm{\nu}$ is a vector adding up arbitrary elements in $\mathrm{NN}$ and $\mathrm{NNN}$, the notation $l+\nu$ represents a unique site label associated with the position $x_l \mathbf{x} + y_l \mathbf{y} + \bm{\nu}$.
The first three lines of Eq.~\ref{je} define the kinetic energy current $\mathbf{J}_K$,
and the fourth to the seventh lines correspond to the potential energy current $\mathbf{J}_P$.
The heat current is $\mathbf{J}_Q=\mathbf{J}_E-\mu \mathbf{J}$.

Fourier transforming the fermion operators 
\begin{equation}
    c^\dagger_{l,\sigma}=\frac{1}{\sqrt{N}}\sum_{\mathbf{k}}e^{-i\mathbf{k}\cdot\mathbf{R}_l}c^\dagger_{\mathbf{k},\sigma},
\end{equation}
where $N$ is the number of sites,
we can transform Eq.~\ref{j}, and the first three lines of Eq.~\ref{je}~($\mathbf{J}_K$), which yield
\begin{eqnarray}
    \mathbf{J} = \sum_{\mathbf{k},\sigma}  \mathbf{v}_{\mathbf{k}} c_{\mathbf{k},\sigma}^\dagger c_{\mathbf{k},\sigma},\\
    \mathbf{J}_{K} = \sum_{\mathbf{k},\sigma} \epsilon_{\mathbf{k}} \mathbf{v}_{\mathbf{k}} c_{\mathbf{k},\sigma}^\dagger c_{\mathbf{k},\sigma},
\end{eqnarray}
as summations in $\mathbf{k}$-space.
Here, $\epsilon_{\mathbf{k}}$ is the band energy at momentum $\mathbf{k}$, determined by the $U=0$ non-interacting limit of the Hamiltonian in Eq.~\ref{hubbard}, $\hat{H}(U=0)=\sum\limits_{\mathbf{k},\sigma}\epsilon_{\mathbf{k}}c_{\mathbf{k},\sigma}^\dagger c_{\mathbf{k},\sigma}$, and $\mathbf{v}_{\mathbf{k}} \equiv \partial \epsilon_{\mathbf{k}}/\partial\mathbf{k}$ is the band velocity.

The two-particle (four-fermion) contribution to $\mathbf{J}_P$ selects charge flow involving a process where an electron flows from/to a site already occupied by an electron of opposite spin. 
In other words, nonzero contributions to the two-particle contribution to $\mathbf{J}_{P}$ rely on breaking and reforming double occupancies, leading to $L_P$ decreasing with increasing doping as double occupancies are reduced.

Using the Kubo formula, the transport coefficients are defined as~\cite{PhysRevB.105.L161103}
\begin{equation}
     L_{O_1O_2}(\omega) = \frac{1}{N \beta} \int_0^\infty d t e^{i(\omega+i0^+) t}\int_0^\beta d\tau\langle{O_1(t-i\tau)O_2(0)}\rangle, \label{definitioncoefficient}
\end{equation}
where $O_1$ and $O_2$ are Hermitian operators that can be chosen as any one of the current operators introduced previously, and $\beta=1/T$ is the inverse temperature.
Here, $t$ is real-time, without confusion with the hopping matrix elements in the Hamiltonian, and 
\begin{equation}
O_1(t-i\tau) = e^{i(\hat{H}-\mu \hat{N})(t-i\tau)} O_1 e^{-i(\hat{H}-\mu \hat{N})(t-i\tau)},
\end{equation}
where $\hat{N}$ is the operator for the total number of particles in the system.
For Hamiltonians such as Eq.~\ref{hubbard}, one can show $L_{O_1O_2}(\omega)=L_{O_2O_1}(\omega)$ from Eq.~\ref{definitioncoefficient} for any operators $O_1$, $O_2 \in \{J_{x}$, $J_{Q,x}$, $J_{K,x}$,  $J_{P,x}\}$~\cite{Shastry_2009}.
We consider $O_1=O_2=O$, and set $Z=\mathrm{Tr}(e^{-\beta(\hat{H}-\mu \hat{N})})$ as the partition function.
From Eq.~\ref{definitioncoefficient}, we obtain
\begin{equation}
     \Re L_{OO}(\omega) =  \frac{\pi}{ZN\beta\omega} \sum\limits_{i_1,i_2}|\langle {i_1|O|i_2}\rangle|^2  e^{-\beta E_{i_1}}\left(1-e^{-\beta\omega}\right) \delta(\omega+E_{i_1}-E_{i_2}),
     \label{definitioncoefficient_useful}
\end{equation}
where $|i_1\rangle$ ($E_{i_1}$) are eigenstates (eigenvalues) of $\hat{H}-\mu \hat{N}$.
We use DQMC to measure the correlation functions in imaginary time 
\begin{eqnarray}
\langle{T_\tau O(\tau)O(0)}\rangle
&\equiv& \frac{1}{Z}\mathrm{Tr}\left(e^{-\beta(\hat{H}-\mu \hat{N})} T_\tau e^{\tau (\hat{H}-\mu \hat{N})}O e^{-\tau (\hat{H}-\mu \hat{N}) } O\right) \nonumber \\
&=&\frac{1}{Z}\sum\limits_{i_1,i_2} |\langle{i_1|O|i_2}\rangle|^2 e^{-\beta E_{i_1}}e^{\tau(E_{i_1}-E_{i_2})}. \label{imaginarytimecorr}
\end{eqnarray}
We then apply MaxEnt~\cite{jarrell1996bayesian,PhysRevB.82.165125} to $\langle{T_\tau O(\tau)O(0)}\rangle$ and obtain $\Re L_{OO}(\omega)$~\cite{PhysRevB.105.L161103,edwin} using the relation
\begin{equation}
\frac{1}{N\beta}\langle{T_\tau O(\tau)O(0)}\rangle =  \int_0^{\infty} d\omega \Re L_{OO}(\omega) \frac{\omega\cosh [\omega(\tau-\beta/2) ]}{\pi\sinh [\beta\omega/2 ]},
\end{equation}
which can be derived from Eqs.~\ref{definitioncoefficient_useful} and \ref{imaginarytimecorr}.
$\Re L_{OO}(\omega)$ is guaranteed to be positive definite in Eq.~\ref{definitioncoefficient_useful}.
In this work, we measure $\langle{T_\tau O(\tau)O(0)}\rangle$ and apply MaxEnt for $O \in \{J_{x}$, $J_{Q,x}$, $J_{K,x}$,  $J_{P',x}$,  $\lambda_Q J_{Q,x}+J_{x}$, $\lambda_K J_{K,x}+J_{x}$,  $\lambda_{P'} J_{P',x}+J_{x}\}$.
Here, ``$x$'' is the $x$ direction and $J_{P',x}$ is defined as $J_{P,x}-\mu J_{x}$. $\lambda_Q$, $\lambda_K$ and $\lambda_{P'}$ are non-zero real constants.

 \begin{figure}
    \centering
    \includegraphics[width=\textwidth]{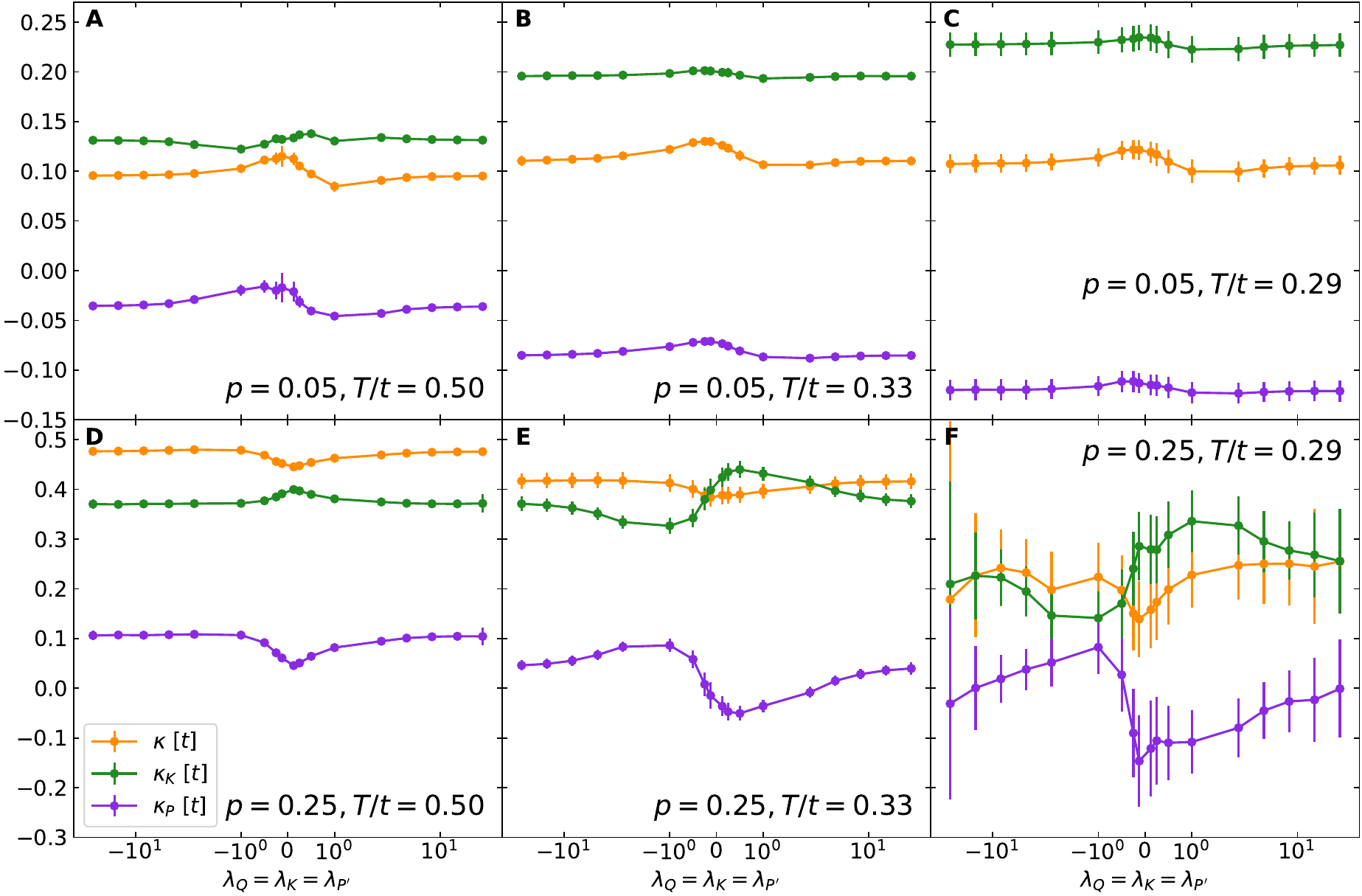}
    \caption{
    $\lambda_Q$, $\lambda_K$, and $\lambda_{P'}$-dependence of $\kappa$, $\kappa_K$ and $\kappa_P$ for a few sets of different doping $p$ and temperature $T$.
     Here, $\lambda_Q$, $\lambda_K$, and $\lambda_{P'}$ are kept the same.
     For MaxEnt analytic continuation, a flat model function is used for all parameters.  
Parameters: $U/t=8$ and $t'/t=-0.25$.
    }
    \label{fig:lambdadependence}
\end{figure}

The kinetic/potential contribution to the DC longitudinal thermal conductivity under the condition of zero electrical current is~\cite{PhysRevB.105.L161103}
\begin{equation}
 \kappa_{K/P} = \beta^2 \left(L_{J_{K/P,x}J_{Q,x}} - \frac{L_{J_{K/P,x}J_x}L_{J_xJ_{Q,x}}}{L_{J_xJ_x}}\right), \label{kappa}
\end{equation}
where the DC transport coefficient $L_{O_1O_2}$ is the $\omega = 0$ value of Eq.~\ref{definitioncoefficient} and is purely real.
Here, for $\kappa_{P}$, we transform Eq.~\ref{kappa} to
\begin{equation}
 \kappa_{P} = \beta^2 \left(L_{J_{P',x}J_{Q,x}} - \frac{L_{J_{P',x}J_x}L_{J_xJ_{Q,x}}}{L_{J_xJ_x}}\right). \label{kappaPanother}
\end{equation}
Even though the expressions are equivalent, using $J_{P',x}$ instead of $J_{P,x}$ in the relevant transport coefficients is a strategy to reduce error propagation.
There are multiple ways to combine correlation functions to obtain the same quantity, as seen in Eq.~\ref{kappa} and Eq.~\ref{kappaPanother}.
Different choices in real computations can result in different magnitudes for the final statistical error after error propagation.
For example, subtracting large quantities that yield a small result can lead to large statistical errors relative to the difference. To reduce the statistical error, we obtain $\kappa$ via
 \begin{equation}
     \kappa = \beta^2 \left(L_{J_{Q,x}J_{Q,x}} - \frac{L_{J_{Q,x}J_x}^2}{L_{J_xJ_x}}\right), \label{eq:kappawhole}
 \end{equation}
where $L_{J_{Q,x}J_x}$ is calculated from~\cite{PhysRevB.95.121104} 
\begin{eqnarray}
L_{J_{Q,x}J_x} =  \left(L_{(\lambda_Q J_{Q,x}+J_{x})(\lambda_Q J_{Q,x}+J_{x})} - \lambda_Q^2 L_{J_{Q,x}J_{Q,x}} - L_{J_{x}J_{x}}\right)/(2 \lambda_Q). \label{lambdaQresult}
\end{eqnarray}
We define and calculate $\kappa_{K,0}$ and $\kappa_{P',0}$ as
\begin{eqnarray}
    \kappa_{K,0} =  \beta^2 \left(L_{J_{K,x}J_{K,x}} - \frac{L_{J_{K,x}J_x}^2}{L_{J_xJ_x}}\right), \label{K0eq} \\
     \kappa_{P',0} =  \beta^2 \left(L_{J_{P',x}J_{P',x}} - \frac{L_{J_{P',x}J_x}^2}{L_{J_xJ_x}}\right),
     \label{kappap0}
\end{eqnarray}
where $L_{J_{K,x}J_x}$ and $L_{J_{P',x}J_x}$ are calculated from
\begin{eqnarray}
L_{J_{K,x}J_x}  = (L_{(\lambda_K J_{K,x}+J_{x})(\lambda_K J_{K,x}+J_{x})} - \lambda_K^2 L_{J_{K,x}J_{K,x}} - L_{J_{x}J_{x}})/(2 \lambda_K),  \label{lambdaKresult} \\ 
L_{J_{P',x}J_x} = (L_{(\lambda_{P'} J_{P',x}+J_{x})(\lambda_{P'} J_{P',x}+J_{x})} - \lambda_{P'}^2 L_{J_{P',x}J_{P',x}} - L_{J_{x}J_{x}})/(2 \lambda_{P'}). \label{lambdaPresult}
\end{eqnarray}
Then from Eqs.~\ref{kappa}, \ref{kappaPanother}, \ref{eq:kappawhole}, \ref{K0eq}, and \ref{kappap0}, we have
\begin{eqnarray}
    \kappa_{K} =  \kappa_{K,0} + (\kappa-\kappa_{K,0}-\kappa_{P',0})/2, \\ 
    \kappa_{P} =  \kappa_{P',0}  + (\kappa-\kappa_{K,0}-\kappa_{P',0})/2,
\end{eqnarray}
which are used for the calculation of $\kappa_K$ and $\kappa_P$.
If all transport coefficients $L_{O_1O_2}$ are exact, then results for $\kappa$, $\kappa_K$ and $\kappa_P$ are independent of the choices for $\lambda_Q$, $\lambda_K$, and $\lambda_{P'}$. 
However, systematic and statistical errors in $L_{O_1O_2}$ propagate to $\kappa$, $\kappa_K$ and $\kappa_P$, so they exhibit some degree of $\lambda_Q$, $\lambda_K$, and $\lambda_{P'}$ dependence (Fig.~\ref{fig:lambdadependence}). Here $\lambda\equiv\lambda_Q=\lambda_K=\lambda_{P'}$.
As long as $\lambda\gtrsim 1$, the $\lambda$ dependence is relatively weak. 
Therefore, as a reasonable choice, $\lambda_Q=\lambda_K=\lambda_{P'}=2$ is used in this work, except for $T/t\geq 4$, where $\lambda_Q=\lambda_K=\lambda_{P'}=0.5$ is used.
This is because for high temperatures, the chemical potential $\mu$ in $J_{Q,x}=J_{E,x}-\mu J_x$ and $J_{P',x}=J_{P,x}-\mu J_x$ has a large magnitude, and $L_{J_{Q,x}J_{Q,x}}$ and $L_{J_{P',x}J_{P',x}}$ are much larger than $L_{J_{x}J_{x}}$.
Therefore, relatively small $\lambda_Q=\lambda_K=\lambda_{P'}=0.5$ were chosen here to reduce error propagation through Eqs.~\ref{lambdaQresult} and \ref{lambdaPresult}.

\begin{figure}
    \centering
    \includegraphics[width=\textwidth]{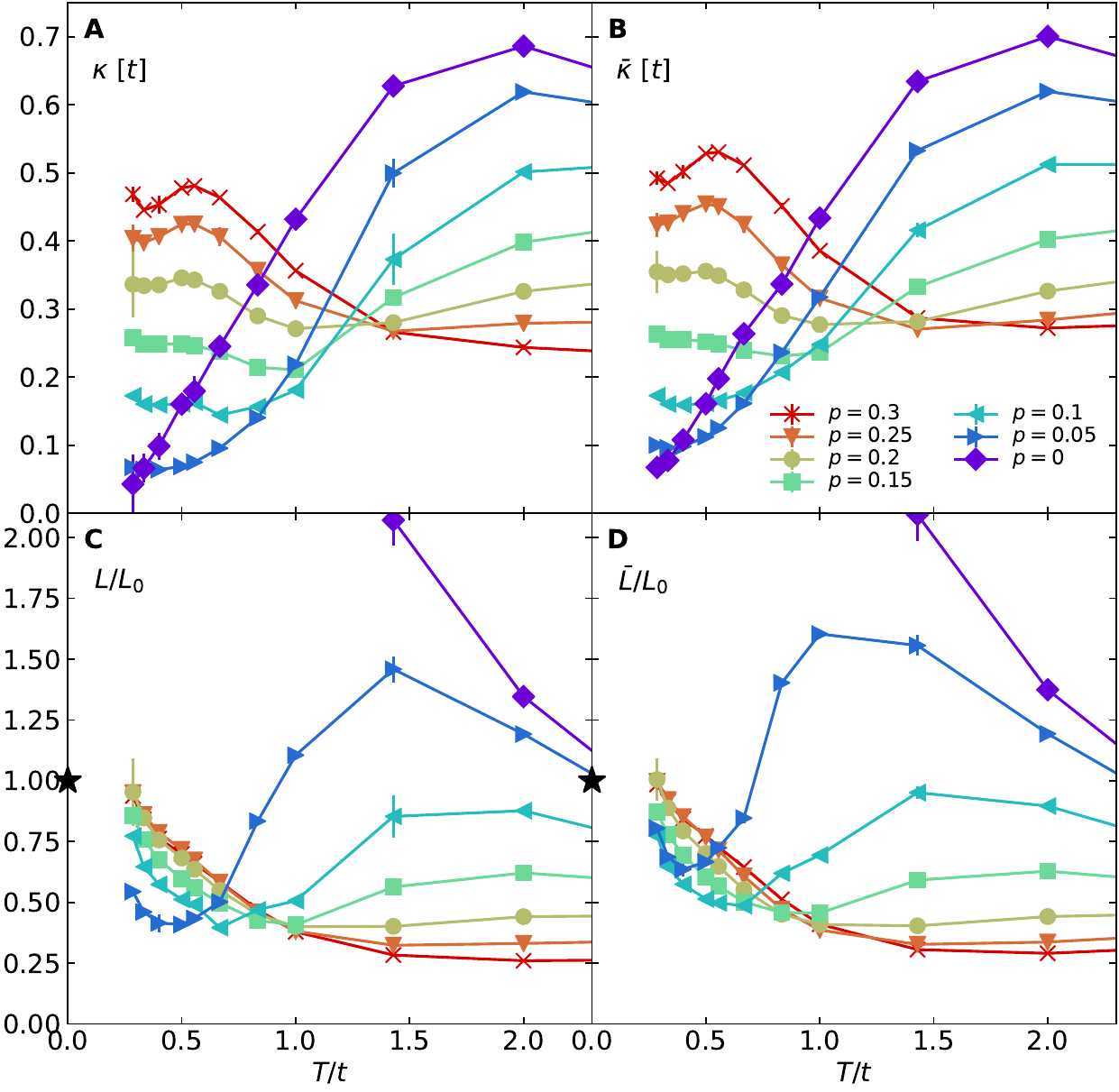}
    \caption{
    (\textbf{A}) DC thermal conductivity $\kappa$ for zero charge current, the same as Fig.~1(\textbf{A}) in the main text.
    (\textbf{B})
    DC thermal conductivity $\bar{\kappa}$ for zero electric field.
    (\textbf{C})
    Lorenz number $L\equiv\kappa/{T\sigma}$ normalized by $L_0\equiv\pi^2/3$, the same as Fig.~2 in the main text.
    (\textbf{D})
    Lorenz number $\bar{L}\equiv\bar{\kappa}/{T\sigma}$, also normalized by $L_0$.
    Parameters: $U/t=8$ and $t'/t=-0.25$.
    }
    \label{fig:twokappa}
\end{figure}

The thermal conductivity $\kappa$ defined under the condition of zero charge current (Eq.~\ref{eq:kappawhole}) can be compared to the
thermal conductivity $\bar{\kappa}$ measured under the condition of zero electric field~\cite{Shastry_2009,PhysRevB.105.L161103}, where
\begin{equation}
    \bar{\kappa} = \beta^2 L_{J_{Q,x}J_{Q,x}}.
\end{equation}
Results for $\kappa$ and $\bar{\kappa}$ are shown in Figs.~\ref{fig:twokappa}(\textbf{A}) and (\textbf{B}), respectively, while
Figs.~\ref{fig:twokappa}(\textbf{C}) and (\textbf{D}) plot the respective Lorenz numbers $L\equiv\kappa/(T\sigma)$ and $\bar{L}\equiv\bar{\kappa}/(T\sigma)$.
Using a representative parameter set $U/t=8$ and $t'/t=-0.25$ as an example, for the low-temperature regime shown in Fig.~\ref{fig:twokappa}, $\kappa$ and $\bar{\kappa}$, as well as $L$ and $\bar{L}$, show similar temperature and doping dependence and have similar magnitudes. Therefore the results and conclusions for the Wiedemann-Franz ratio hold for both quantities.

\begin{figure}
    \centering
    \includegraphics[width=0.85\textwidth]{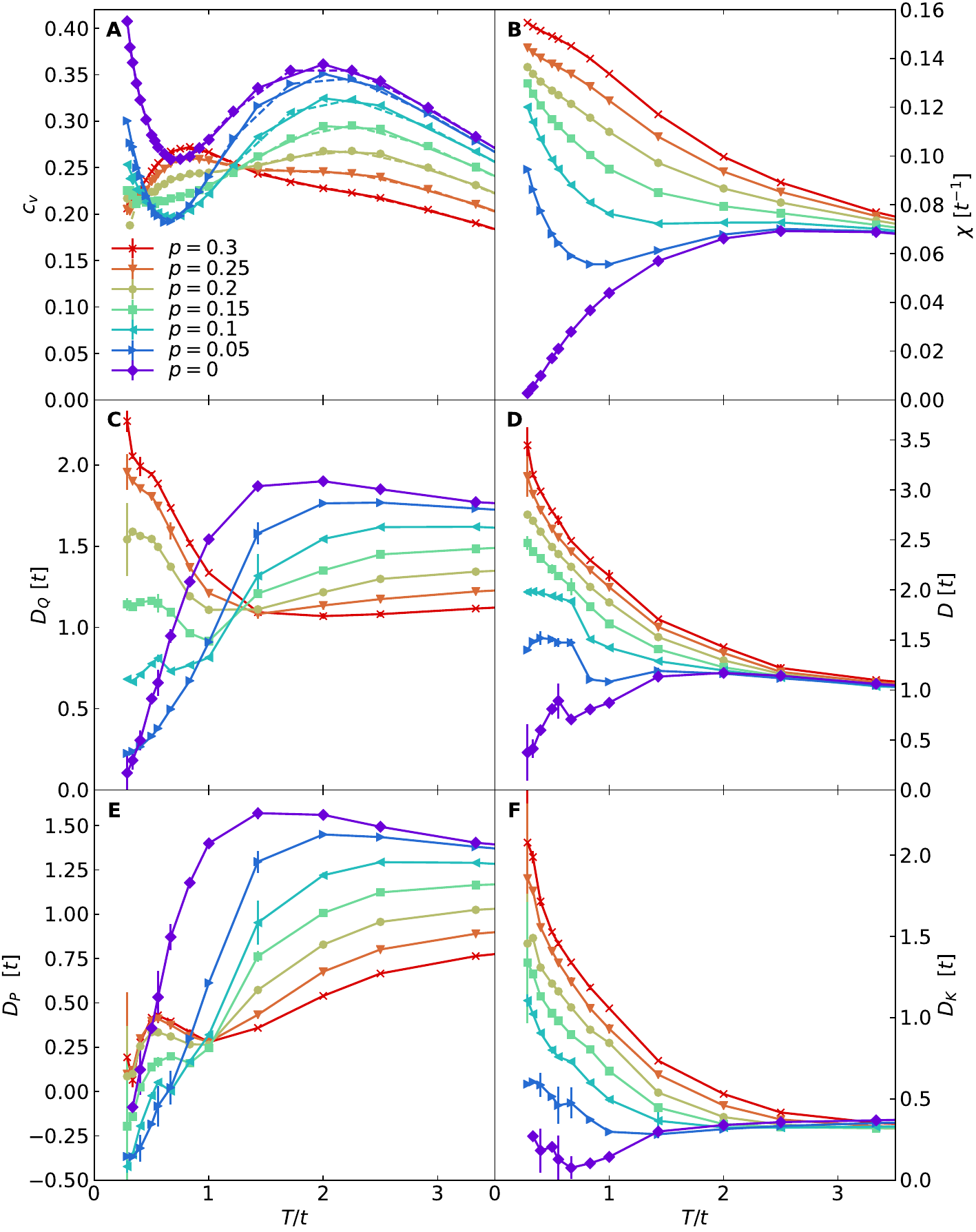}
    \caption{
    (\textbf{A}) Specific heat $c_v$.
    Solid lines are results from the fluctuation method and dashed lines are the results from the finite difference method.
    (\textbf{B}) Charge compressibility $\chi$.
    (\textbf{C}) Thermal diffusivity $D_Q$.
    (\textbf{D}) Charge diffusivity $D$.
    (\textbf{E}) and (\textbf{F}) show the decomposition of thermal diffusivity $D_Q$ into potential $D_P$ and kinetic $D_K$ contributions, respectively.
    Error bars in (\textbf{A}) and (\textbf{B}) denote standard error of the mean determined by jackknife resampling.
    Error propagation is used for $c_v$ obtained from the finite difference method.
    Parameters: $U/t=8$ and $t'/t=-0.25$.
    }
    \label{fig:cchi}
\end{figure}

As discussed in Ref.~\cite{PhysRevB.105.L161103}, the specific heat $c_v$
may be obtained from the average energy $\langle \hat{H} \rangle$ at different temperatures for fixed densities, directly calculating $\delta (\langle \hat{H} \rangle/N)/\delta T$ by choosing a reasonable finite temperature interval $\delta T$, or from the energy fluctuations via 
\begin{equation}
    c_v = \frac{\beta}{N} \left(\Lambda_{\hat{H} \hat{H}} - \frac{\Lambda_{\hat{H} \hat{N}}^2}{\Lambda_{\hat{N}\hat{N}}}\right).
    \label{specificheatformula}
\end{equation}
Here, $\Lambda_{O_1 O_2} = \beta(\langle{O_1 O_2}\rangle-\langle{O_1}\rangle\langle{O_2}\rangle)$. Similarly, the charge compressibility $\chi$ can be calculated using the fluctuation method $\chi=\Lambda_{\hat{N}\hat{N}}/N$. 
Figure~\ref{fig:cchi} shows the DQMC results for (\textbf{A}) the specific heat $c_v$ and (\textbf{B}) the charge compressibility $\chi$~\cite{edwin}.
Note that in Fig.~\ref{fig:cchi}(\textbf{A}) the results for $c_v$ are consistent between the finite difference and fluctuation methods.

\begin{figure}
    \centering
    \includegraphics[width=0.52\textwidth]{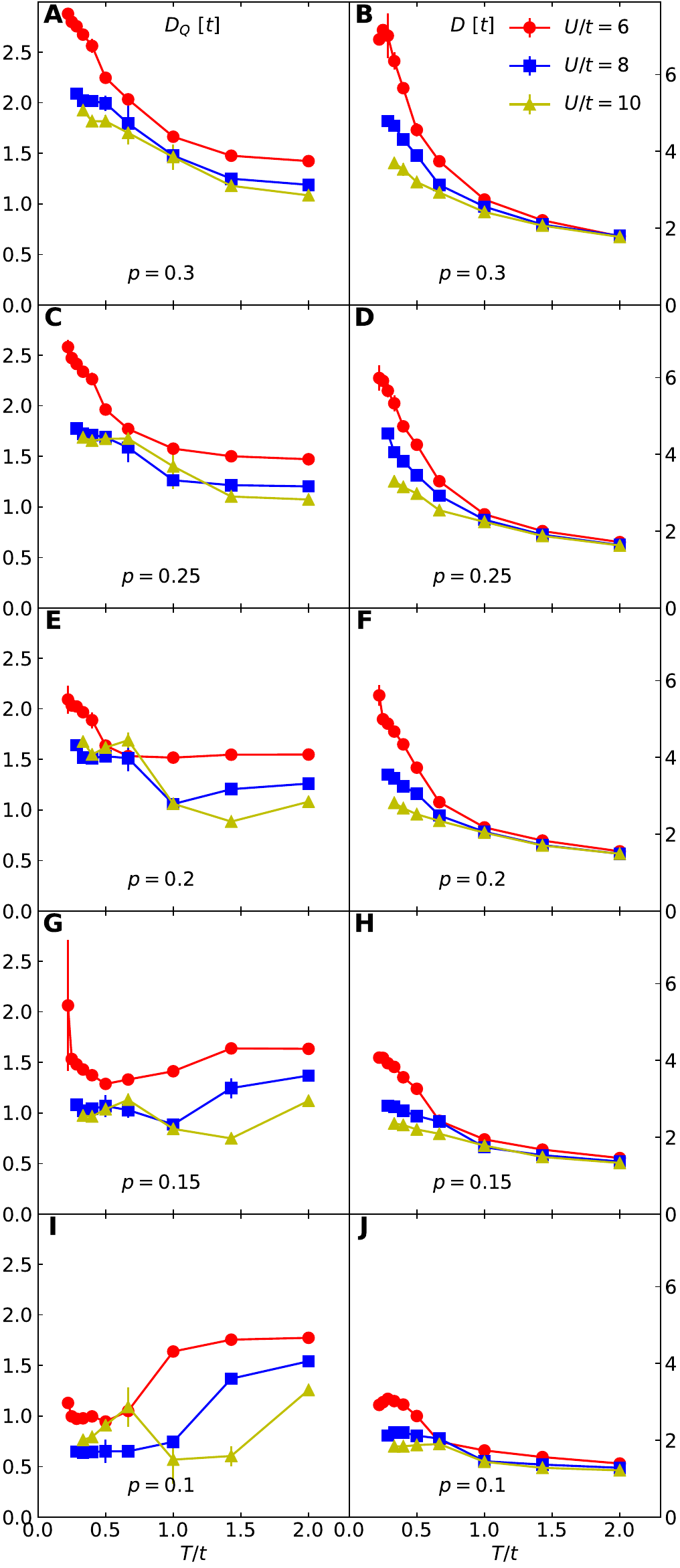}
    \caption{
    Hubbard $U$ dependence of $D_Q$ and $D$ for $t'/t=0$.
    First column is $D_Q$ and the second column is $D$.
    }
    \label{fig:udeptp0}
\end{figure}

\begin{figure}
    \centering
    \includegraphics[width=0.52\textwidth]{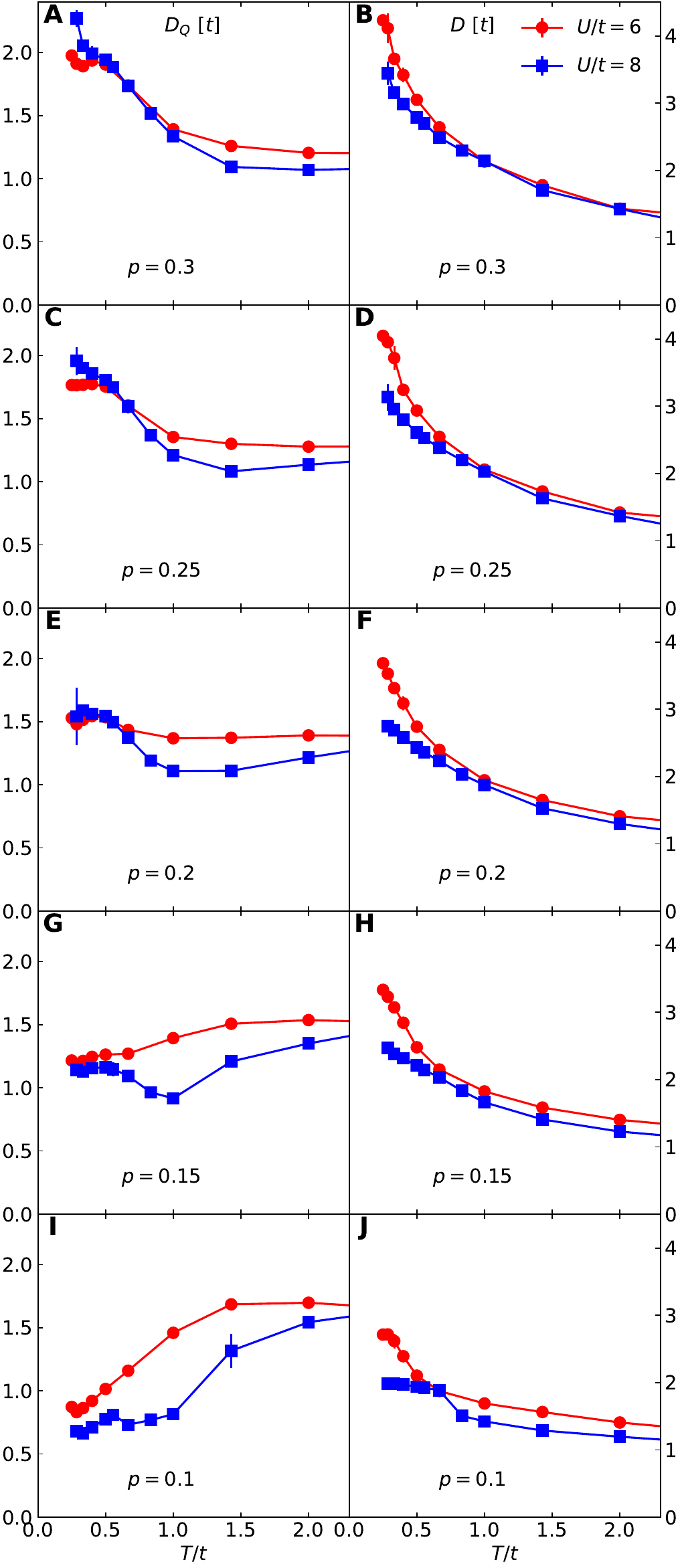}
    \caption{
 Hubbard $U$ dependence of $D_Q$ and $D$ similar to Fig.~\ref{fig:udeptp0} but for $t'/t=-0.25$.
    }
    \label{fig:udeptp025}
\end{figure}

The Einstein relations may be used to calculate the thermal diffusivity $D_Q\equiv\kappa/c_v$ and charge diffusivity $D\equiv\sigma/\chi$~\cite{edwin,PhysRevLett.122.186601}, shown in Figs.~\ref{fig:cchi}(\textbf{C}) and (\textbf{D}).
(The fluctuation method is used for calculating both $c_v$ and $\chi$ when determining both $D_Q$ and $D$.) 
There is a cross-over temperature where the doping dependence of $D_Q$ changes, similar to $\kappa$;
and in the low temperature regime below $T_{\mathrm{xo}}$, $D_Q$ behaves similarly to $D$ with respect to doping and temperature dependence, as well as changes in the value of Hubbard $U$ (see Figs.~\ref{fig:udeptp0} and \ref{fig:udeptp025}).

\begin{figure}
    \centering
    \includegraphics[width=\textwidth]{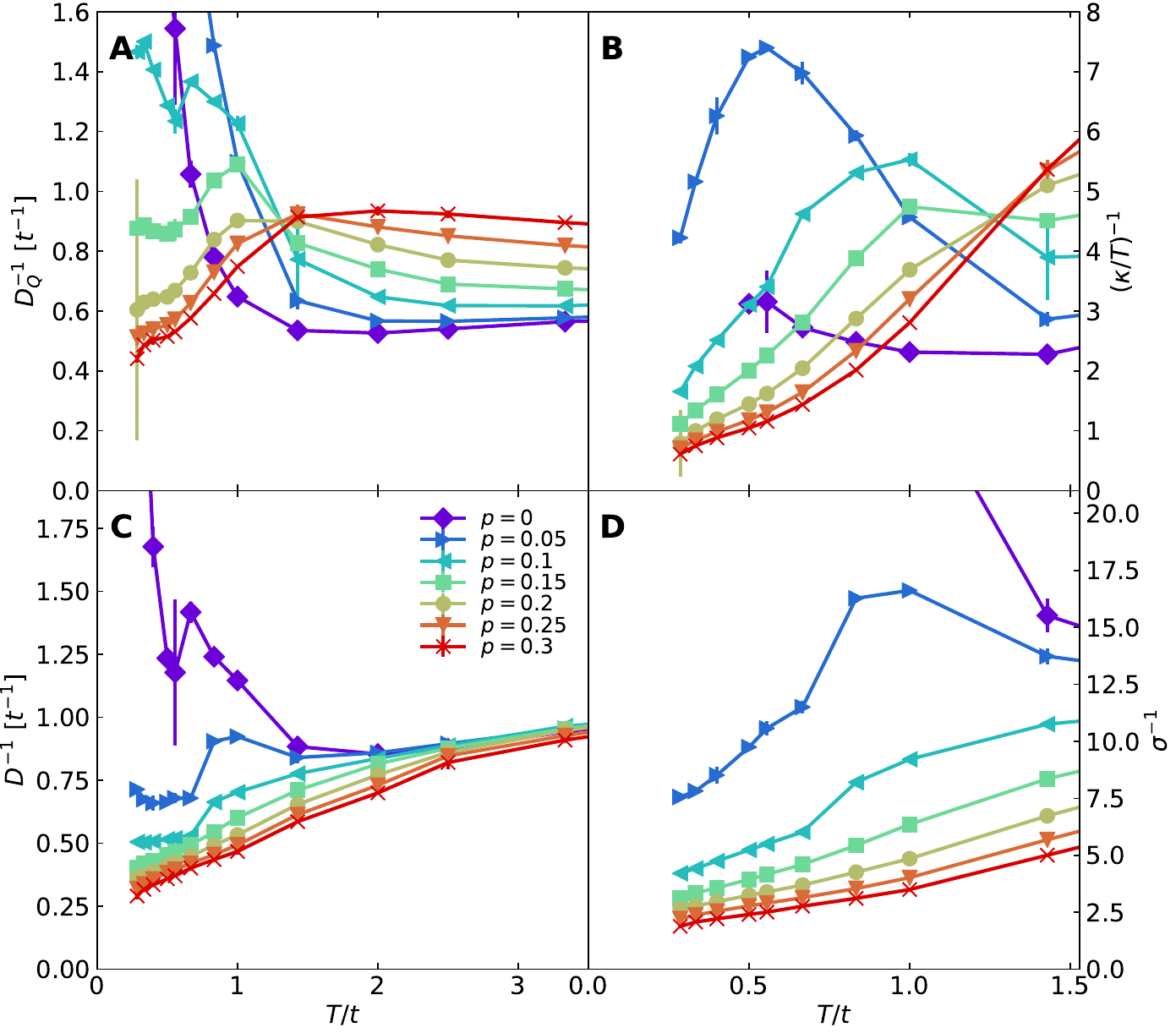}
    \caption{
    Inverse $D_Q$, $\kappa/T$, $D$ and $\sigma$. 
    Parameters: $U/t=8$ and $t'/t=-0.25$.
    }
    \label{fig:inv}
\end{figure}

The potential-kinetic separation of the thermal diffusivity, defined as $D_P\equiv\kappa_P/c_v$ and $D_K\equiv\kappa_K/c_v$, is shown in Figs.~\ref{fig:cchi}(\textbf{E}) and (\textbf{F}), respectively. 
The inverses $D_Q^{-1}$, $(\kappa/T)^{-1}$, $D^{-1}$, and $\sigma^{-1}$ in Fig.~\ref{fig:inv} reveal strange metallic $\sim T$ behavior in the doped metallic regime, consistent with earlier results~\cite{edwin}.

\begin{figure}
    \centering
    \includegraphics[width=\textwidth]{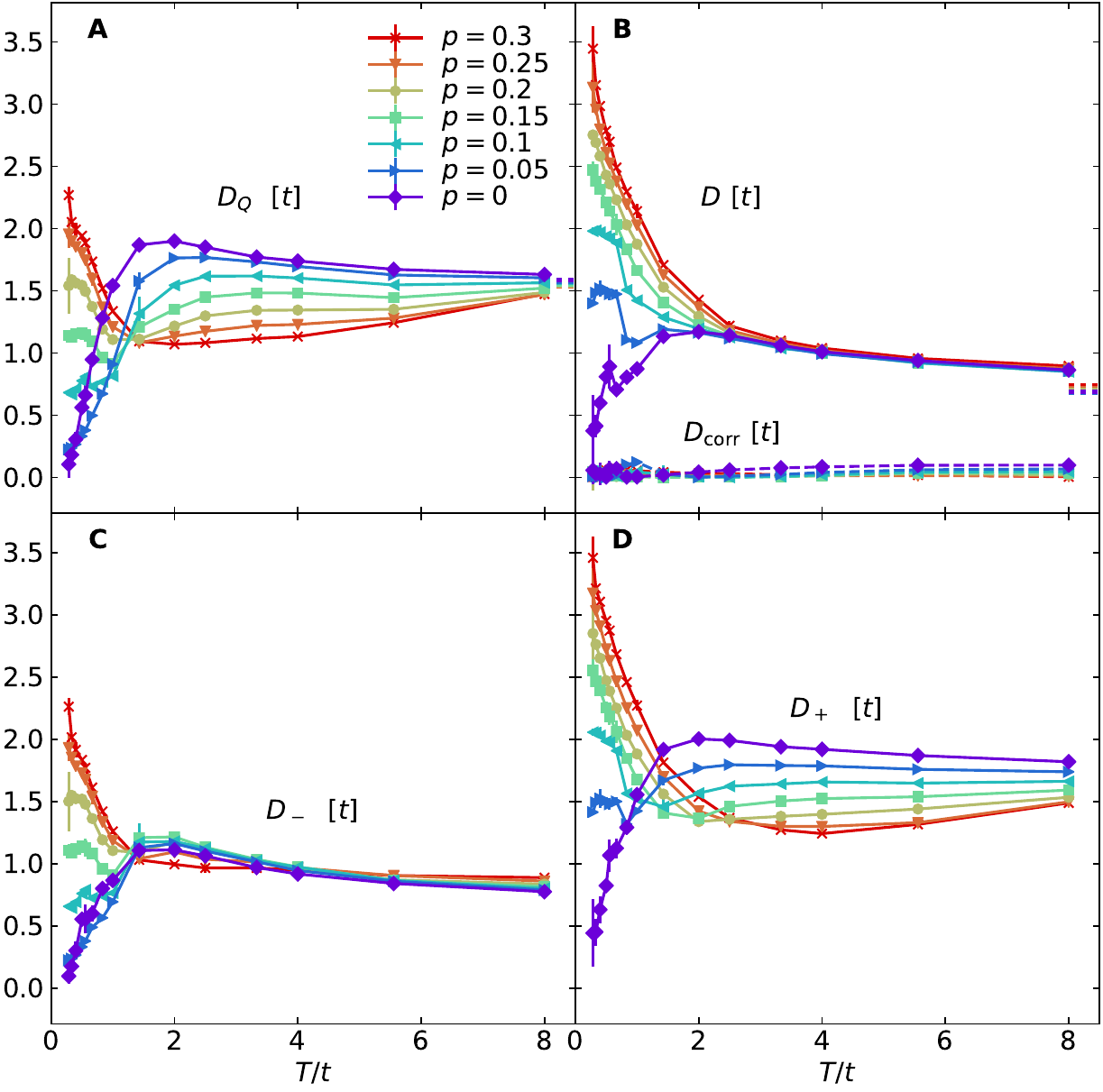}
    \caption{
     Comparisons between $D_Q$, $D$, $D_{\mathrm{corr}}$ and $D_\pm$ over a large temperature regime.
    (\textbf{A}) Thermal diffusivity $D_Q$.
    (\textbf{B}) Charge diffusivity $D$ (solid lines), and correction term $D_{\mathrm{corr}}$ (dashed lines).
    (\textbf{C}) $D_-$.
    (\textbf{D}) $D_+$.
    Dotted lines in (\textbf{A}) and (\textbf{B}) are infinite-temperature limits obtained by the moments expansion method for $\kappa$ and $\sigma$, and analytic calculation for $c_v$ and $\chi$, which will be described later for Fig.~\ref{fig:hight}.
    Parameters: $U/t=8$ and $t'/t=-0.25$.
    }
    \label{fig:dcorr}
\end{figure}

$D_Q$ and $D$ are not strictly independent diffusivities in the heat and charge channels due to non-zero thermoelectric effects in our model.
From Refs.~\cite{hartnoll2015theory,PhysRevLett.122.186601}, $D_{\pm}$ are determined by
\begin{eqnarray}
    D_+ D_- &=& DD_Q \label{d1}\\
    D_+ + D_- &=& D+D_Q +D_{\mathrm{corr}}, \label{d2}
\end{eqnarray}
where $D_{\mathrm{corr}}$ is the correction term defined in Refs.~\cite{hartnoll2015theory,PhysRevLett.122.186601}. 
Fig.~\ref{fig:dcorr} shows that $D_{\mathrm{corr}}$ is orders of magnitude smaller than either $D_Q$ or $D$.
Therefore, $D_{\pm} \approx D$ or $D_Q$ depending on their relative magnitudes. 
At high temperatures, $D_Q>D$ and $D_+\approx D_Q$, as shown in Figs.~\ref{fig:dcorr}(\textbf{A}) and (\textbf{D}). At low temperatures, $D>D_Q$ and $D_+\approx D$, shown in  Figs.~\ref{fig:dcorr}(\textbf{B}) and (\textbf{D}). 
$D_-$ in Fig.~\ref{fig:dcorr}(\textbf{C}) takes the smaller value between $D_Q$ and $D$.
In the high-temperature limit, $D_+$ is about twice that of $D_-$, consistent with Ref.~\cite{PhysRevLett.122.186601}.
Therefore, in our parameter regime, corrections $D_{\mathrm{corr}}$ due to thermoelectric effects are negligible compared with $D_Q$ and $D$, so that $D_Q$ and $D$ are good approximations to the independent diffusivities $D_{\pm}$. 

\noindent\underline{Model Function Dependence}

\begin{figure}
    \centering
    \includegraphics[width=0.9\textwidth]{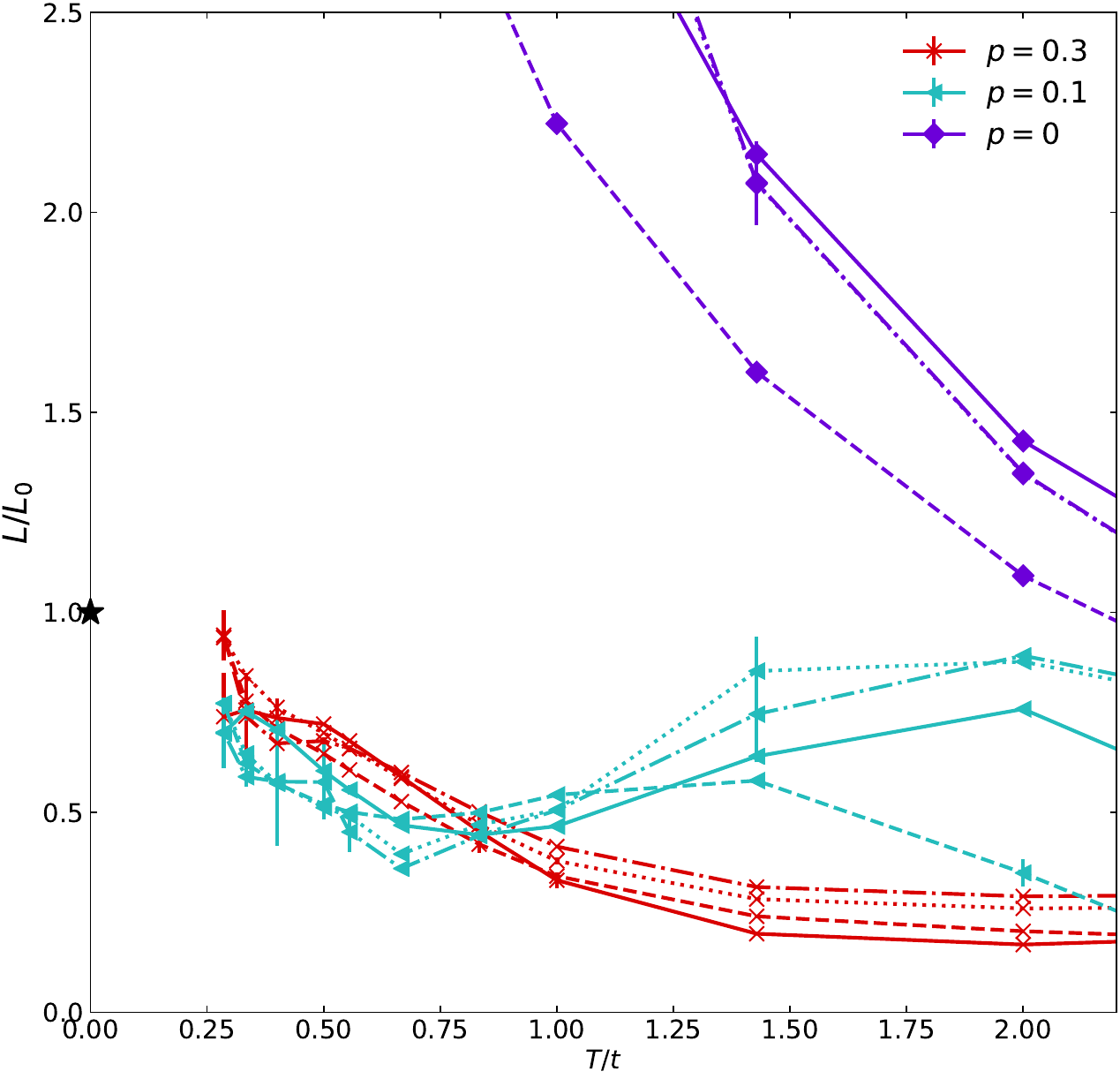}
    \caption{
Lorenz number $L$ obtained using 
method 1 (dotted lines), method 2 (dashed lines), method 3 (dashed dotted lines), and method 4 (solid lines). 
See the ``Model Function Dependence'' subsection for definitions of these methods.
For $p=0$ in the displayed range, the dotted line appears hidden as its values are very close to those on the dashed dotted line.
Parameters: $U/t=8$ and $t'/t=-0.25$.
    }
    \label{fig:differentana}
\end{figure}

Model function dependence of the results for the Lorenz number is shown in Fig.~\ref{fig:differentana}.
Four methods of constructing the model function are analyzed here: 
\begin{itemize}
\item Method 1: The annealing procedure from high to low temperatures, which is used in the main text. 
\item Method 2: Annealing from low to high temperatures by starting the procedure using results from method 1 at the lowest temperature as the initial model function. 
\item Method 3: Using the model functions constructed from the infinite-temperature-limit spectra but with changing chemical potentials for all temperatures. 
\item Method 4: Using a flat model function for all temperatures.
\end{itemize}
Fig.~\ref{fig:differentana} shows MaxEnt results obtained using the four methods with qualitatively similar behavior. Therefore, conclusions and the discussion in the main text remain the same regardless of how one constructs the model function in these four distinct ways.

\noindent\underline{Miscellaneous Supplementary Data}

In this subsection a number of checks are presented to support the conclusions given in the main text.

\begin{figure}
    \centering
    \includegraphics[width=0.9\textwidth]{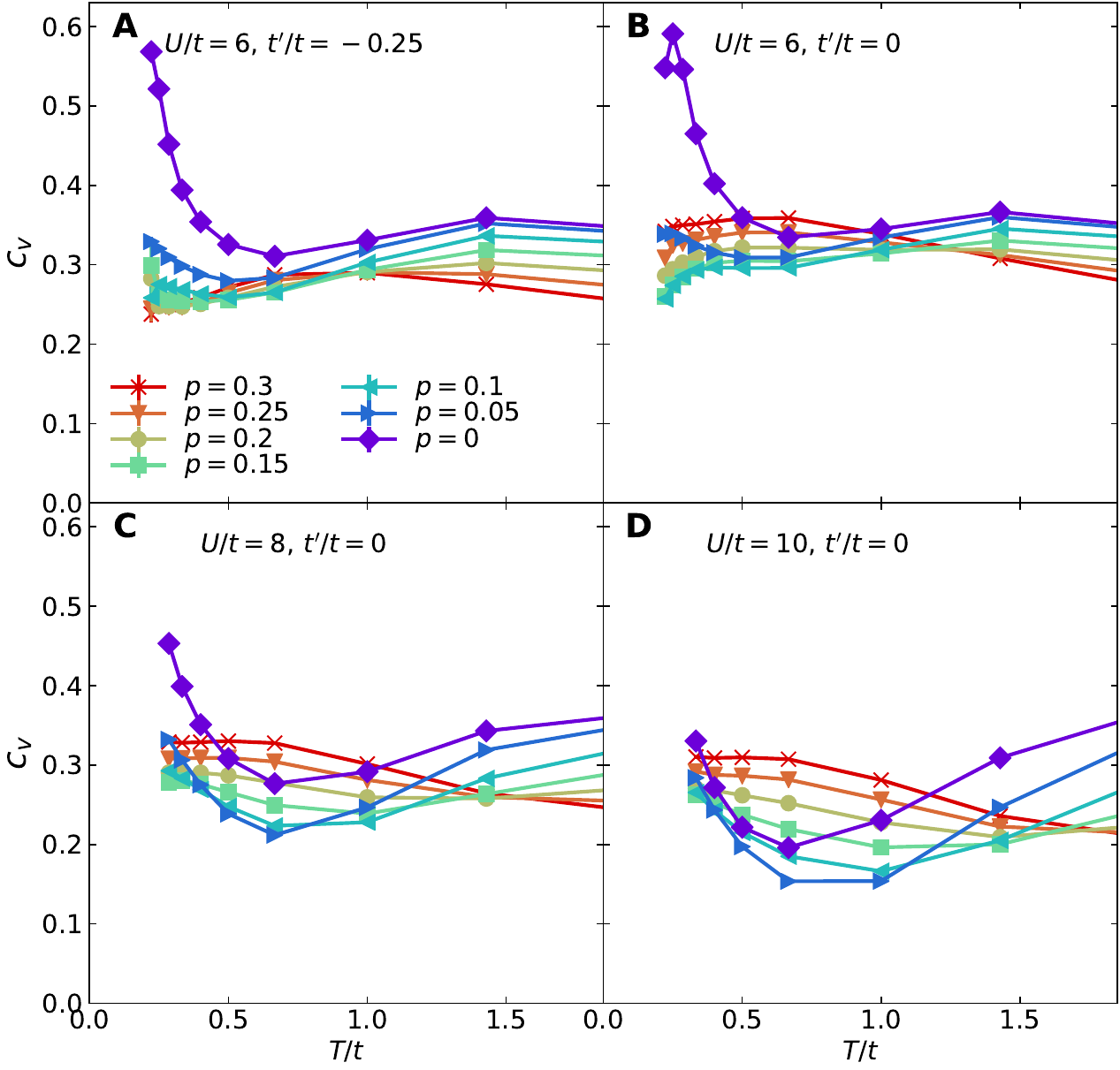}
    \caption{
Specific heat $c_v$ obtained via the fluctuation method Eq.~\ref{specificheatformula} for 
    (\textbf{A}) $U/t=6$ and $t'/t=-0.25$;
    (\textbf{B}) $U/t=6$ and $t'/t=0$;
    (\textbf{C}) $U/t=8$ and $t'/t=0$;
    (\textbf{D}) $U/t=10$ and $t'/t=0$.
    }
    \label{fig:cs}
\end{figure}

\begin{figure}
    \centering
    \includegraphics[width=0.9\textwidth]{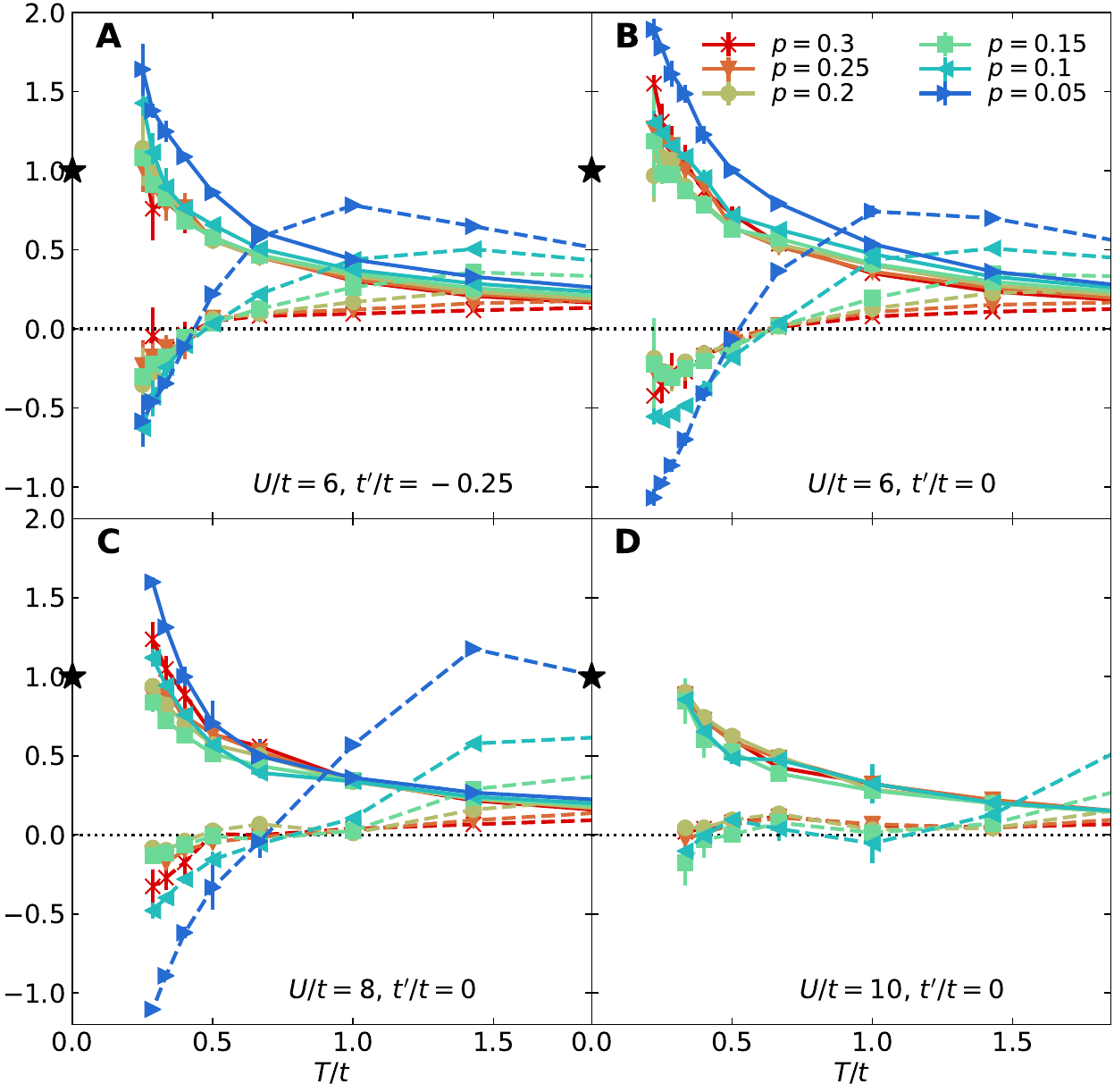}
    \caption{
Kinetic/potential decomposition of $L/L_0$ for  (\textbf{A}) $U/t=6$ and $t'/t=-0.25$;
    (\textbf{B}) $U/t=6$ and $t'/t=0$;
    (\textbf{C}) $U/t=8$ and $t'/t=0$;
    (\textbf{D}) $U/t=10$ and $t'/t=0$.
Solid lines are $L_K/L_0$ and dashed lines are $L_P/L_0$.
   The black stars mark the value $1$.
   The dotted lines mark the value $0$.
    }
    \label{fig:separations}
\end{figure}

Fig.~\ref{fig:cs} demonstrates the parameter dependence of the specific heat $c_v$. The high-temperature peak position of $c_v$ is controlled by the energy scale set by $U$, similar to that of $L$ in Fig.~3 in the main text.

Fig.~\ref{fig:separations} plots the kinetic/potential decomposition of $L/L_0$ for different parameters. For all parameter choices, the kinetic/potential components exhibit behavior similar to Fig.~4 in the main text.

\begin{figure}
    \centering
    \includegraphics[width=0.85\textwidth]{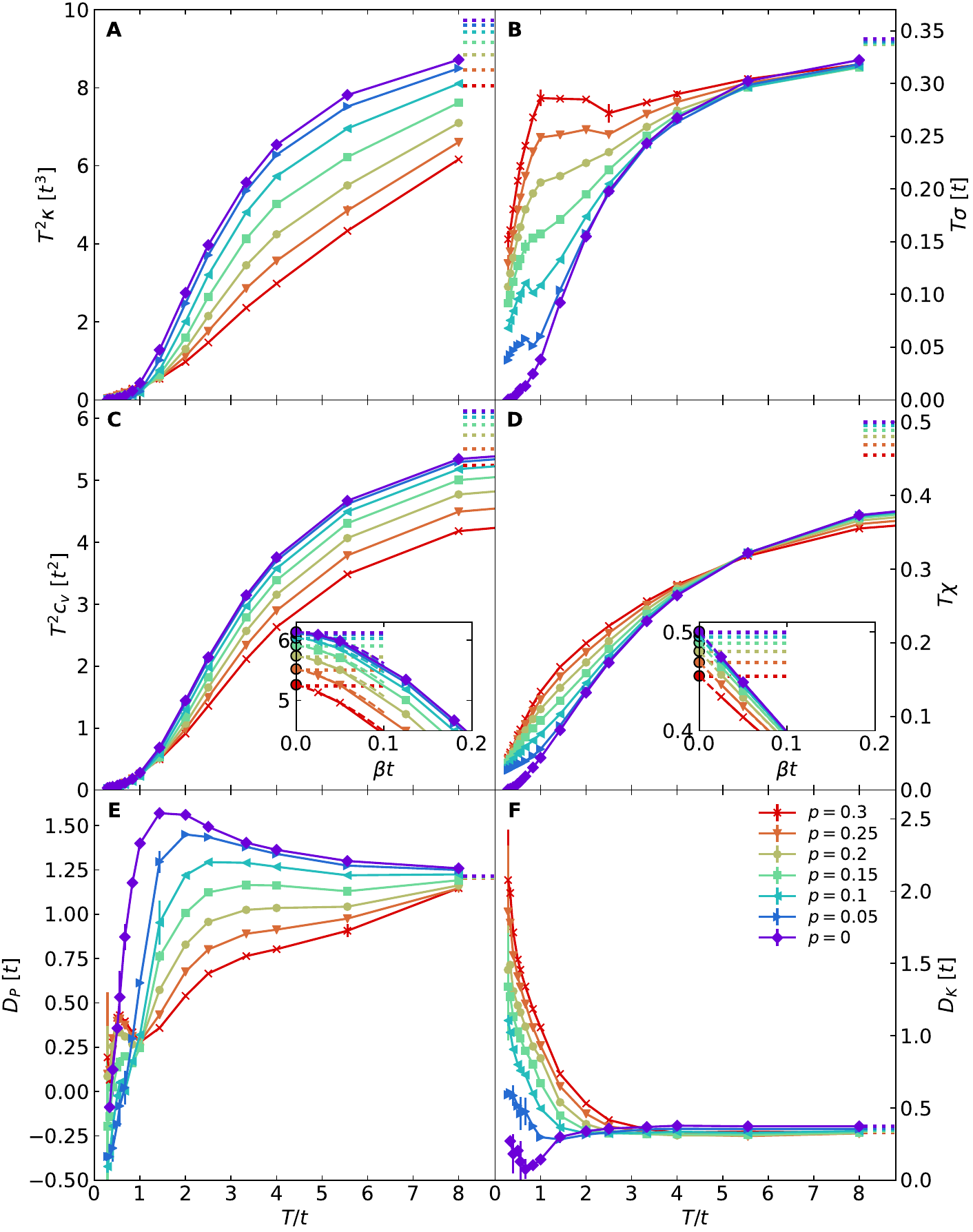}
    \caption{Behavior over a large range of temperatures for (\textbf{A}) $T^2\kappa$, (\textbf{B}) $T \sigma $ (same data as in Fig.~1(\textbf{C}) in the main text), (\textbf{C}) $T^2c_v$ (using the fluctuation method), (\textbf{D}) $T\chi$, (\textbf{E}) $D_P$, and (\textbf{F}) $D_K$,
    and a comparison to the infinite-temperature limits for each quantity.
Insets for (\textbf{C}) and (\textbf{D}) are $T^2 c_v$ and $T\chi$ versus $\beta t$, respectively, and include data for temperatures $T/t \ge 8$.
In these insets, dashed lines mark the cubic spline extrapolated function; filled circles mark the extrapolated positions for $\beta t=0$.
The dotted lines in (\textbf{A}) to (\textbf{F}), including the insets, are infinite-temperature limits.
Error bars for $T^2 c_v$ and $T\chi$ are from jackknife resampling.
Parameters: $U/t=8$ and $t'/t=-0.25$.
}
    \label{fig:hight}
\end{figure}

Fig.~\ref{fig:hight} shows measurements of several quantities up to $T/t=8$ and compares them to the corresponding infinite-temperature limits.
All measurements approach their corresponding infinite-temperature limits as temperature increases.
The infinite-temperature limits of the transport properties were calculated using the moments expansion method.
The infinite-temperature limits of thermodynamic quantities $c_v$ and $\chi$, given by fluctuations $\Lambda_{O_1 O_2}$, were calculated analytically. When $T\rightarrow \infty$, $e^{-\beta \hat{H}}=1$, so
$\langle{O_1 O_2}\rangle = \mathrm{Tr}(e^{\beta \mu\hat{N}} O_1 O_2)/\mathrm{Tr}(e^{\beta \mu\hat{N}})$ and $\langle{O_1}\rangle = \mathrm{Tr}(e^{\beta \mu\hat{N}}  O_1)/\mathrm{Tr}(e^{\beta \mu\hat{N}}).$
The calculation of the right-hand-side traces is straightforward in the occupation basis, when $O_1$ and $O_2$ are expressed in fermion operators $\mathit{c}_{l,\mathit{\sigma}}^{\dagger}$ and $\mathit{c}_{l,\mathit{\sigma}}$.

\noindent\underline{Trotter Error Analysis} \label{sec:Trotter}

\begin{figure*}
    \centering
    \includegraphics[width=0.8\textwidth]{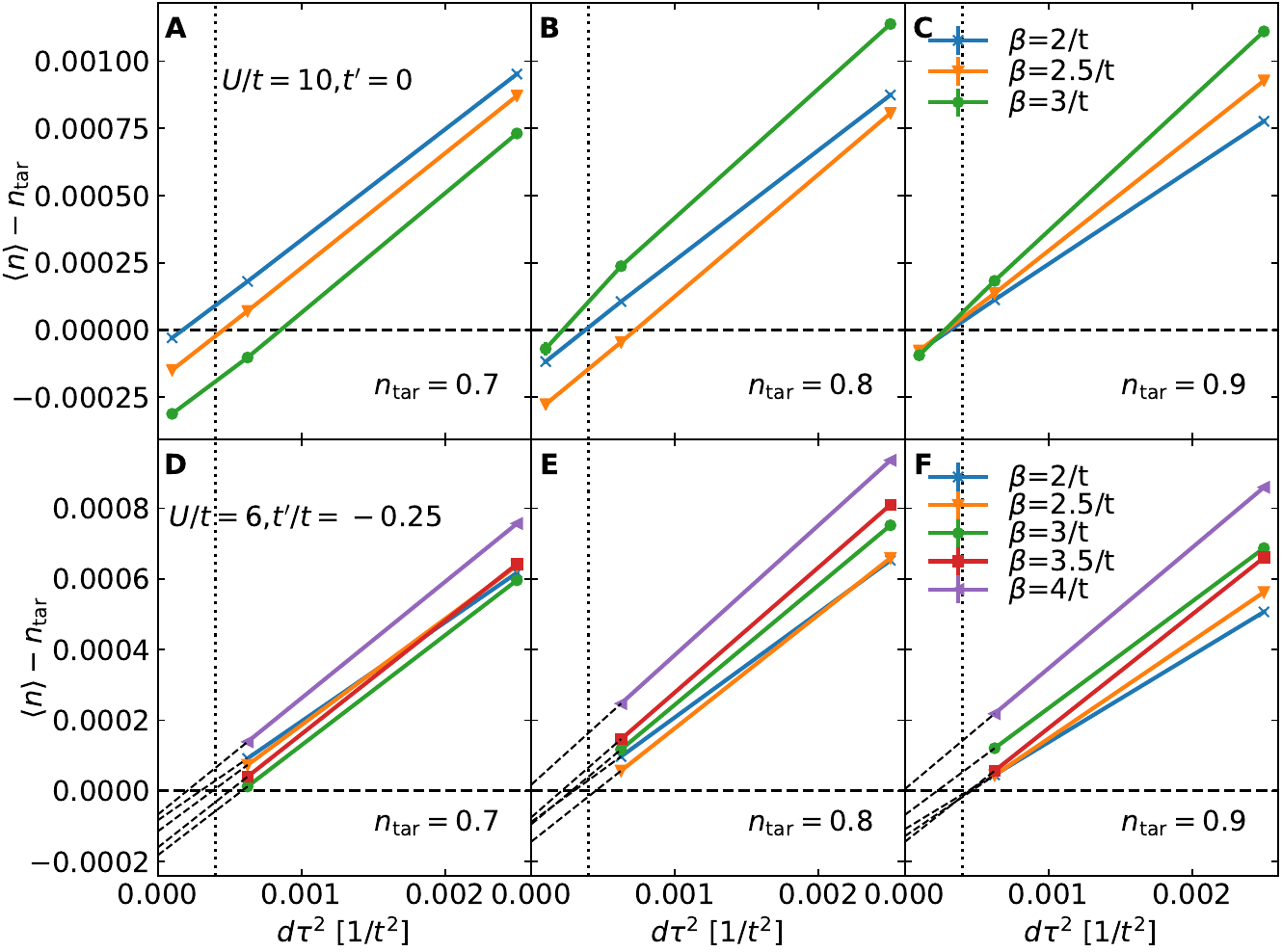}
    \caption{Trotter error analysis for chemical potential tuning. Difference between measured density $\langle n \rangle$ and target density $n_{\mathrm{tar}}$ for (\textbf{A})-(\textbf{C}) $U/t=10$, $t'/t=0$ and (\textbf{D})-(\textbf{F}) $U/t=6$, $t'/t=-0.25$.
    Vertical dotted lines mark $d\tau=0.02/t$, which is the discretization interval used for chemical potential tuning.
    The $\mu$ tuned for $n_{\mathrm{tar}}$ at $d\tau=0.02/t$ is used to measure $\langle n \rangle$ for varying values of $d\tau$ on the same curve. 
    Thin dashed lines in (\textbf{D})-(\textbf{F}) are straight-line extrapolations as a guide for eye. 
    Thick dashed lines indicate $\langle n \rangle - n_{\mathrm{tar}}=0$.
    Error bars, which are smaller than the data points, are $\pm 1$ standard error of the mean determined by jackknife resampling.
    }
    \label{fig:densitytrotter}
\end{figure*}

Trotter error enters measurements of thermodynamic and transport properties in two ways.
First, it enters in measurements of density $\langle n \rangle$ during chemical potential tuning.
After the chemical potential $\mu$ is determined for each target density, Trotter error enters in the measurements themselves.

For chemical potential tuning,
Fig.~\ref{fig:densitytrotter} shows the difference between the measured density $\langle n \rangle$ and the target density $n_{\mathrm{tar}}$ as a function of discretization interval $d\tau$, using $\mu$ values obtained from tuning with $d\tau=0.02/t$. 
Fig.~\ref{fig:densitytrotter}(\textbf{A})-(\textbf{C}) shows that the Trotter error scales as $\sim  d\tau^2$.
Extrapolating $\langle n \rangle- n_{\mathrm{tar}}$ to $d\tau^2 = 0$ indicates an estimation of the ``true'' value of $\langle n \rangle(d\tau=0)- n_{\mathrm{tar}}$, i.e. the systematic error of density due to finite $d\tau=0.02/t$.
For $U/t=10,\,t'/t=0$ (the largest $U$ considered in this paper), and $U/t=6,\,t'/t=-0.25$ (lower temperatures are achievable), the estimated deviation $|\langle n \rangle(d\tau=0) - n_{\mathrm{tar}}|$ is within $10^{-4}$ as shown in Fig.~\ref{fig:densitytrotter}, which represents an upper bound of Trotter error in density for the parameters considered in this paper.

\begin{figure*}
    \centering
    \includegraphics[width=\textwidth]{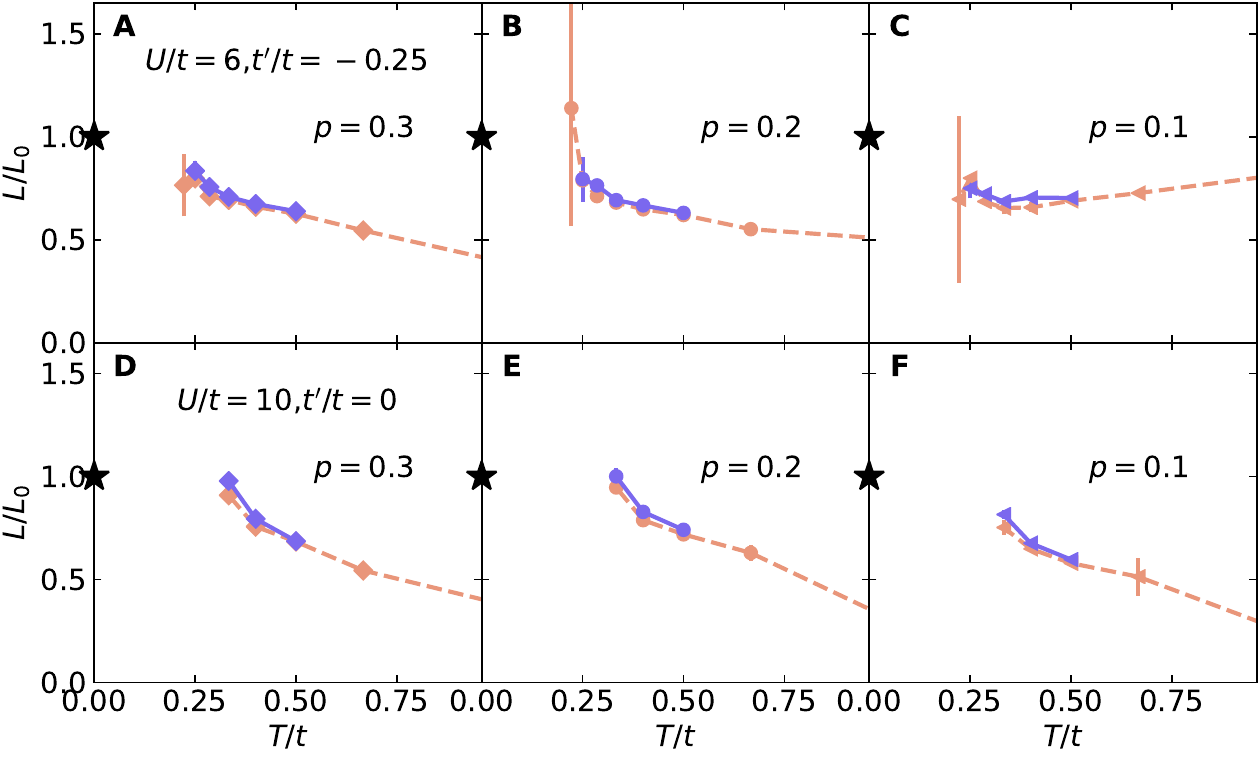}
    \caption{Trotter error analysis of $L$ for (\textbf{A})-(\textbf{C}) $U/t=6,\,t'/t=-0.25$, and (\textbf{D})-(\textbf{F}) $U/t=10,\,t'/t=0$. Dashed lines are for $d\tau = 0.05/t$ and solid lines are for $d\tau = 0.025/t$.
    }
    \label{fig:trotterL}
\end{figure*}

We show the Trotter error for $L$ in Fig.~\ref{fig:trotterL}, for representative parameters $U/t=6,\, t'/t=-0.25$ [(\textbf{A})-(\textbf{C})] and  $U/t=10,\, t'/t=0$ [(\textbf{D})-(\textbf{F})].
Results obtained with $ d\tau=0.05/t$ and $0.025/t$ show minimal difference.
For transport measurements, in addition to direct changes in the Trotter error, changing $d\tau$ may affect analytic continuation, as the number of imaginary time points changes for fixed inverse temperature $\beta$.
Analyzing Fig.~\ref{fig:trotterL},
we conclude that $d\tau=0.05/t$ is small enough to prevent Trotter error from affecting our conclusions, and is also a reasonable value for stable MaxEnt analytic continuation. 

\noindent\underline{Finite Size Effects Analysis}

\begin{figure*}
    \centering
    \includegraphics[width=\textwidth]{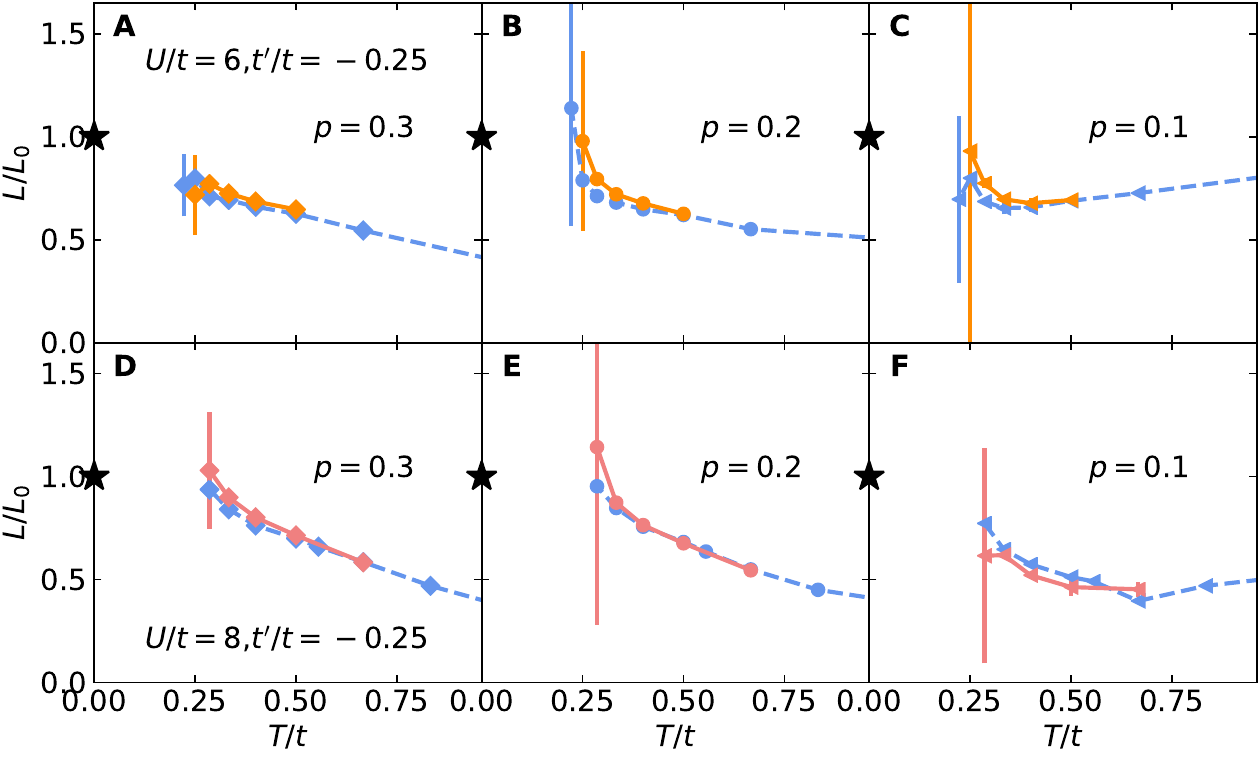}
    \caption{
    Finite size dependence of $L$ with $t'/t=-0.25$, for (\textbf{A})-(\textbf{C}) $U/t=6$, and (\textbf{D})-(\textbf{F}) $U/t=8$.
    Dashed lines are obtained on clusters of size $8\times 8$. Solid lines are obtained on clusters of size $12\times 12$ for (\textbf{A})-(\textbf{C}) and $10\times 10$ for (\textbf{D})-(\textbf{F}).}
    \label{fig:finitesize_L}
\end{figure*}

\begin{figure*}
    \centering
    \includegraphics[width=\textwidth]{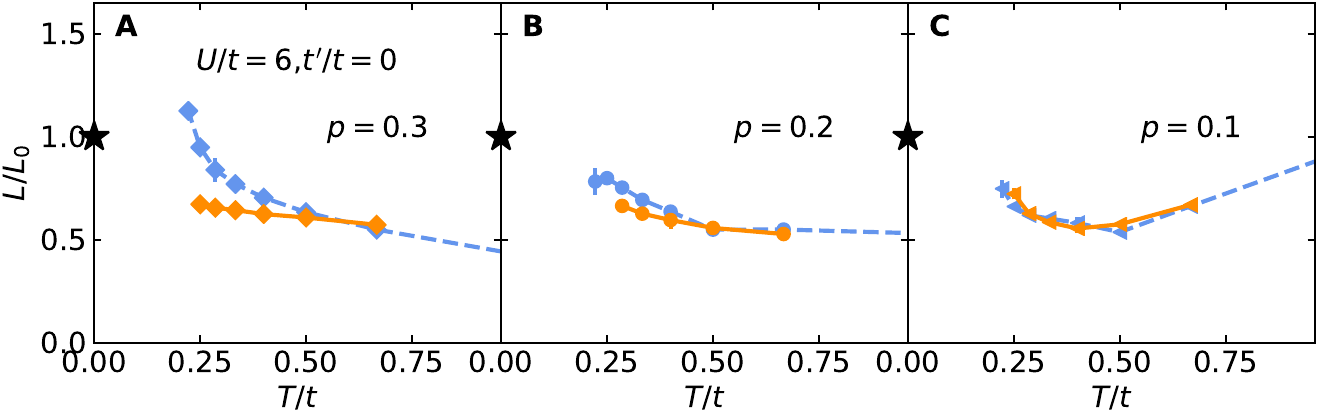}
    \caption{
    Finite size analysis of $L$ with $t'/t=0$.
     Dashed lines are obtained on clusters of size $8\times 8$. Solid lines are obtained on clusters of size $12\times 12$.
     Simulation parameters: $U/t=6$, $t'/t=0$.
    }
    \label{fig:finitesize_L_tp0}
\end{figure*}

In Fig.~\ref{fig:finitesize_L}, we compare results of $L$ on $8\times 8$ and $12\times 12$ clusters for $U/t=6, t'/t=-0.25$, and on $8\times 8$ and $10\times 10$ clusters for $U/t=8,\,t'/t=-0.25$.
The differences between the results are minimal and do not affect our conclusions.
A similar finite size analysis for $U/t=6, t'/t=0$ is shown in Fig.~\ref{fig:finitesize_L_tp0}.
We observe larger finite size effect for $t'/t=0$ compared with $t'/t=-0.25$, and for higher doping and lower temperature,
because of sharper Drude peaks and more delocalized nature of the system. 
Lattice size slightly changes the low temperature behaviors of $L$ for $t'/t=0$, but the overall conclusions are not significantly affected.
Smaller $U$ generally should also cause larger finite size effects as the system becomes more delocalized.
Thus, our analysis up to $30\%$ doping, down to $U/t=6$, including both $t'/t=-0.25$ and $t'=0$, and down to the lowest accessible temperatures roughly represent an upper bound on the finite size effects, 
given the parameters considered in this manuscript.
\end{document}